\documentclass[a4paper,11pt]{article}

\pdfoutput=1 

\usepackage{jheppub} 

\usepackage{slashed}
\usepackage{subcaption}
\usepackage{xcolor}
\usepackage{cleveref}
\usepackage[T1]{fontenc} 
\usepackage[utf8]{inputenc}
\usepackage{mathtools}

\DeclarePairedDelimiter\floor{\lfloor}{\rfloor}

\newcommand{\cA}{\mathcal{A}}
\newcommand{\cM}{\mathcal{M}}

\newcommand{\wa}{ w \cdot \mathfrak{a} }

\newcommand{\crossBox}{\,\rotatebox{90}{\scalebox{0.5}[0.9]{$\bowtie$}}}

\makeatletter
\newsavebox{\@brx}
\newcommand{\llangle}[1][]{\savebox{\@brx}{\(\m@th{#1\langle}\)}%
  \mathopen{\copy\@brx\kern-0.5\wd\@brx\usebox{\@brx}}}
\newcommand{\rrangle}[1][]{\savebox{\@brx}{\(\m@th{#1\rangle}\)}%
  \mathclose{\copy\@brx\kern-0.5\wd\@brx\usebox{\@brx}}}
\makeatother

\subheader{\normalsize
	\vspace*{-3.7em}
	\begin{flushright}
		CP3-22-15\\
		UUITP-12/22\\
		CALT-TH-2022-009
	\end{flushright}
}

\title{\boldmath Higher-spin scattering in the classical limit}
\title{\boldmath Classical scattering for heavy particles with spin}
\title{\boldmath Searching for Kerr in the 2PM amplitude}

\author[a]{Rafael Aoude,}
\author[b,c]{Kays Haddad,}
\author[d]{and Andreas Helset}
\affiliation[a]{Centre for Cosmology, Particle Physics and Phenomenology (CP3),\\
Universit\'{e} catholique de Louvain, 1348 Louvain-la-Neuve, Belgium}
\affiliation[b]{Department of Physics and Astronomy, Uppsala University, \\
Box 516, 75120 Uppsala, Sweden}
\affiliation[c]{Nordita, Stockholm University and KTH Royal Institute of Technology, \\
Hannes Alfv\'{e}ns v\"{a}g 12, 10691 Stockholm, Sweden}
\affiliation[d]{Walter Burke Institute for Theoretical Physics,
California Institute of Technology,\\ Pasadena, CA 91125, USA}
\emailAdd{rafael.aoude@uclouvain.be}
\emailAdd{kays.haddad@physics.uu.se}
\emailAdd{ahelset@caltech.edu}

\abstract{
The classical scattering of spinning objects is well described by the spinor-helicity formalism for heavy particles.
Using these variables, we derive spurious-pole-free, all-spin opposite-helicity Compton amplitudes (factorizing on physical poles to the minimal, all-spin three-point amplitudes of ref.~\cite{Arkani-Hamed:2017jhn}) in the classical limit for QED, QCD, and gravity.
The cured amplitudes are subject to deformations by contact terms, the vast majority of whose contributions we can fix by imposing a relation between spin structures---motivated by lower spin multipoles of black hole scattering---at the second post-Minkowskian (2PM) order.
For QED and gravity, this leaves a modest number of unfixed coefficients parametrizing contact-term deformations, while the QCD amplitude is uniquely determined.
Our gravitational Compton amplitude allows us to push the state-of-the-art of spinning-2PM scattering to any order in the spin vectors of both objects; we present results here and in the auxiliary file \texttt{2PMSpin8Aux.nb} up to eighth order in the spin vectors.
Interestingly, despite leftover coefficients in the Compton amplitude, imposing the aforementioned relation between spin structures uniquely fixes some higher-spin parts of the 2PM amplitude.
}

\allowdisplaybreaks

\begin{document} 
\maketitle
\flushbottom

\section{Introduction}\label{sec:Introduction}

The observation and analysis of gravitational waves from compact binary coalescence by the LIGO and Virgo collaborations presents a unique lens through which to study gravity and the formation and constitution of dense astrophysical objects \cite{LIGOScientific:2016aoc,LIGOScientific:2017vwq,LIGOScientific:2018mvr,LIGOScientific:2020ibl,LIGOScientific:2021usb}.
Central to these analyses is the comparison to theoretical predictions.
The initial phase of this coalescence, referred to as the inspiral phase, is defined by a large separation between the two bound objects.
In this regime, the gravitational attraction between the bodies is weak, facilitating the computation of observables analytically in an expansion in Newton's constant $G$---the so-called post-Minkowskian (PM) expansion.\footnote{A further expansion in the velocities of the objects leads to the post-Newtonian (PN) expansion; see refs.~\cite{Porto:2016pyg,Levi:2018nxp} for recent reviews.}
Given a framework for converting scattering amplitudes to classical observables, a one-to-one correspondence then exists between the loop expansion of amplitudes describing matter coupled to gravity and the PM expansion of observables.

Indeed, a constantly-growing body of work continues to validate the applicability of scattering-amplitude-based approaches to understanding the inspiral phase of binary coalescence.
The scattering of two Schwarzschild black holes has been understood completely up to 3PM by studying the $2\rightarrow2$ amplitude of two massive scalars \cite{Cheung:2018wkq,Bern:2019nnu,Bern:2019crd,Cheung:2020gyp,Kalin:2020mvi,Kalin:2020fhe,Herrmann:2021lqe,Herrmann:2021tct,DiVecchia:2021bdo,Brandhuber:2021eyq}.
Studies of spinless scattering continue to break new ground, with conservative dynamics at 4PM emerging from the same process evaluated at three-loop order \cite{Bern:2021dqo,Bern:2021yeh,Dlapa:2021npj,Dlapa:2021vgp}.
Tidal effects have also been shown to be accessible to scattering-amplitude techniques, by including contributions from higher-curvature operators to the scattering \cite{Cheung:2020sdj,Haddad:2020que,Kalin:2020lmz,Cheung:2020gbf,Bern:2020uwk,AccettulliHuber:2020dal,Aoude:2020ygw}.

Of course, astrophysical objects carry angular momentum, which must be captured to faithfully describe the dynamics of the inspiral.
A plethora of frameworks have been developed for understanding how to describe such objects in terms of quantum point particles with spin, producing new PM results in the process \cite{Guevara:2017csg,Guevara:2018wpp,Chung:2018kqs,Maybee:2019jus,Guevara:2019fsj,Arkani-Hamed:2019ymq,Chung:2019duq,Damgaard:2019lfh,Aoude:2020onz,Bern:2020buy,Aoude:2020ygw,Liu:2021zxr,Kosmopoulos:2021zoq,Aoude:2021oqj,Jakobsen:2021lvp,Jakobsen:2021zvh,Chen:2021qkk,Jakobsen:2022fcj}.
Up until now, PM dynamics of spinning objects have been computed up to fourth order in the spins of both objects at 2PM \cite{Guevara:2017csg,Guevara:2018wpp,Chung:2018kqs,Damgaard:2019lfh,Bern:2020buy,Kosmopoulos:2021zoq,Chen:2021qkk}, and up to second order in spin at 3PM (including conservative dynamics for general spin orientations and radiative effects when spins are aligned) \cite{Jakobsen:2022fcj}.

Various methods exist for converting scattering amplitudes with or without spin to classical observables, either directly \cite{Kosower:2018adc,Maybee:2019jus,Kalin:2019rwq,Bjerrum-Bohr:2019kec,Cristofoli:2021vyo}, or by first passing through intermediate quantities such as a Hamiltonian \cite{Cheung:2018wkq,Cristofoli:2019neg,Bern:2020buy} or eikonal phase \cite{Amati:1987uf,Guevara:2018wpp,Bern:2020buy,DiVecchia:2021bdo,Bautista:2021wfy,Haddad:2021znf,Adamo:2021rfq}.
Crucially as well, it has been understood how to translate scattering information into quantities pertinent to the motion of bound bodies \cite{Kalin:2019rwq,Kalin:2019inp,Saketh:2021sri,Cho:2021arx}.

\begin{figure}
    \centering
    \includegraphics[scale=0.3]{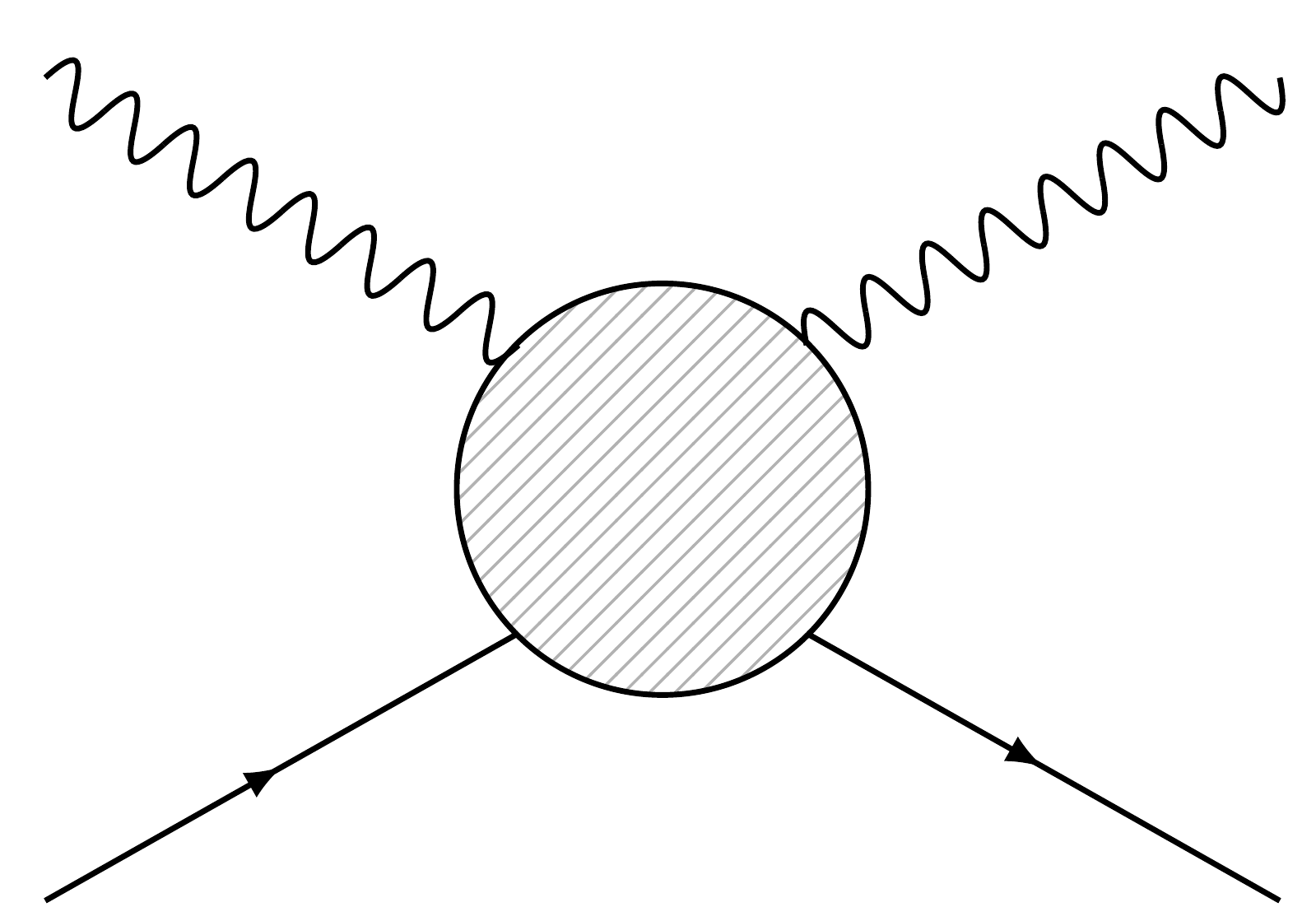}
    \caption{\label{fig:Compton}Compton scattering. The solid lines represent a massive spin-$s$ particle, with the flow of momentum shown by the arrows. The wavy lines represent bosons being emitted from the massive leg.}
\end{figure}

The recent work of ref.~\cite{Chen:2021qkk} presented substantial progress past the 2PM results involving spin that came before, using the Compton amplitude found in ref.~\cite{Arkani-Hamed:2017jhn} to construct the 2PM Hamiltonian up to fourth order in spin.
Before these results can be pushed to the next orders in spin using scattering amplitudes, there is an obstacle that must be overcome.
Constructing classical effects at $n$PM order entails a computation of the classical part of the $2\rightarrow2$ massive scattering amplitude at $(n-1)$-loop level.
Taking a unitarity approach to evaluating the loop amplitude, one needs knowledge of the amplitude for $n$-graviton emission from a massive spinning particle \cite{Neill:2013wsa,Bjerrum-Bohr:2013bxa,Vaidya:2014kza,Bern:2019crd}.
Specifically at $n=2$, classical contributions to the 2PM amplitude are controlled by the opposite-helicity Compton amplitude (see \cref{fig:Compton}) \cite{Neill:2013wsa}.\footnote{This is no longer the case when the same-helicity Compton amplitude is deformed by contact terms or by considering three-point amplitudes that do not describe black holes; see \Cref{sec:2PM} and Appendix B of ref.~\cite{Chen:2021qkk}. We are grateful to Justin Vines for pointing this out.}
If the scattered particles have spins $s_{1,2}\leq2$, current knowledge of the Compton amplitude can be used to obtain 2PM scattering dynamics up to fourth order in the spin vectors, as per ref.~\cite{Chen:2021qkk}.
Above these spins, issues arise.

The Compton amplitude for the emission of two opposite-helicity bosons from a massive, arbitrary spin particle was first constructed in ref.~\cite{Arkani-Hamed:2017jhn}, using Britto-Cachazo-Feng-Witten (BCFW) recursion \cite{Britto:2004ap,Britto:2005fq} on the minimal three-point amplitude in terms of massive spinor-helicity variables. 
It was found to be
\begin{subequations}\label{eq:StandardCompton}
\begin{align}
    \cA_{\text{QED}}(-\mathbf{1}^{s},\mathbf{2}^{s},3^{-},4^{+})&=\frac{y^{2}}{t_{13}t_{14}}\left(\frac{\langle3\mathbf{1}\rangle[4\mathbf{2}]-\langle3\mathbf{2}\rangle[4\mathbf{1}]}{y}\right)^{2s}, \\
    \cM(-\mathbf{1}^{s},\mathbf{2}^{s},3^{-},4^{+})&=\frac{y^{4}}{s_{34}t_{13}t_{14}}\left(\frac{\langle3\mathbf{1}\rangle[4\mathbf{2}]-\langle3\mathbf{2}\rangle[4\mathbf{1}]}{y}\right)^{2s},
\end{align}
in the electromagnetic and gravitational cases respectively.
The QCD case was first presented in ref.~\cite{Johansson:2019dnu}:
\begin{align}
    \cA_{\text{QCD}}(-\mathbf{1}^{s}_{i},\mathbf{2}^{s}_{j},3^{-}_{c},4^{+}_{d})&=\left(\frac{T^{c}_{ik}T^{d}_{kj}}{s_{34}t_{13}}+\frac{T^{d}_{ik}T^{c}_{kj}}{s_{34}t_{14}}\right)y^{2}\left(\frac{\langle3\mathbf{1}\rangle[4\mathbf{2}]-\langle3\mathbf{2}\rangle[4\mathbf{1}]}{y}\right)^{2s}.
\end{align}
\end{subequations}
We have introduced the Mandelstam variables $t_{1i}=(p_{1}-q_{i})^{2}-m^{2}$, $s_{34}=(q_{3}+q_{4})^{2}$, where we use $q_{i}$ to represent massless momenta and $p_{i}$ for massive ones.
We have also abbreviated $y\equiv[4|p_{1}|3\rangle$.
The massive leg has spin $s$.
Fundamental and adjoint color indices are denoted by letters from the middle or beginning of the Latin alphabet, respectively.
The bold notation represents contracting the symmetrized $2s$ little group indices for each massive spinor with an auxiliary variable; see \Cref{sec:OSHPET}.

All amplitudes in \cref{eq:StandardCompton} can be seen to develop unphysical poles at $y=0$ when the spin of the massive particle exceeds the spin of the emitted bosons.
The emergence of this pole coincides with the fact that the product of BCFW shifted three-point amplitudes does not go to zero sufficiently fast as the BCFW shift parameter approaches infinity.
This suggests that BCFW recursion applied at high enough spins requires boundary terms in order to produce well-defined amplitudes.
The same-helicity Compton amplitude was presented for all spins first in ref.~\cite{Johansson:2019dnu}.
It was shown to not possess unphysical poles at any spin, a conclusion that was extended to the amplitude involving $n$ same-helicity photons or gravitons in ref.~\cite{Aoude:2020onz}, and $n$ same-helicity gluons in ref.~\cite{Lazopoulos:2021mna}.

Since the description of classical spinning bodies using amplitudes requires scattering particles with arbitrarily high spin, the lack of a physical opposite-helicity Compton amplitude above spin 2 is problematic.
Several approaches to date have worked towards understanding the opposite-helicity Compton amplitude at high spins \cite{Chung:2018kqs,Falkowski:2020aso,Chiodaroli:2021eug,Bautista:2021wfy}.
Refs.~\cite{Chung:2018kqs,Falkowski:2020aso} constructed boundary terms that removed the unphysical poles without affecting physical factorization channels.
Ref.~\cite{Chiodaroli:2021eug} constructed the spin-$3/2$ gauge theory and spin-$5/2$ gravity Compton amplitudes from a Lagrangian by imposing constraints on a three-point higher-spin current and choosing the best behaved Compton amplitudes in the high-energy limit.
They recovered precisely \cref{eq:StandardCompton} for spins up to 1 and 2 for electromagnetism and gravity, respectively.
At higher spins, however, the results in these three works are not all compatible with one another: refs.~\cite{Falkowski:2020aso,Chiodaroli:2021eug} agree with each other where there is overlap in their results, but not with ref.~\cite{Chung:2018kqs}.
Moreover, our analysis shows that the spin-$3/2$($5/2$) electromagnetic (gravitational) amplitudes of the former two works do not agree with \cref{eq:StandardCompton} at quadratic-(quartic-)order in spin for electromagnetism (gravity).
The all-spin amplitude of ref.~\cite{Chung:2018kqs} disagrees with \cref{eq:StandardCompton} both at cubic and quartic order in spin, as was pointed out in that paper. 
All of these disagreements are attributed to contact terms (parts of the amplitude that vanish on all physical residues), but they are still troublesome because the total spin of the scattered particles should not affect lower spin multipoles in the classical limit if finite-spin amplitudes are to yield reliable classical predictions.
Moreover, the comprehensive results in ref.~\cite{Falkowski:2020aso} are presented recursively with no closed form solution to the recursion, and those of ref.~\cite{Chung:2018kqs} are themselves quite cumbersome and are not easily recast in terms of the classical spin vector.
Finally, ref.~\cite{Bautista:2021wfy} aimed to use solutions from classical general relativity to constrain the Compton amplitude, but the tensor perturbations linked to the gravitational Compton amplitude are not treated there.

In this paper we add to this body of work on the Compton amplitude and improve upon the points listed above.
Our analysis focuses on narrowing down the portion of the Compton amplitude relevant to describing classical black hole scattering at 2PM, which is specifically the leading part of the amplitude in the $\hbar\rightarrow0$ limit.
Towards this end, we recast \cref{eq:StandardCompton} in terms of the heavy on-shell variables of ref.~\cite{Aoude:2020onz}, and study its pole structure for each theory.
Our focus on the $\hbar\rightarrow0$ limit allows us to determine a compact closed form for the spurious-pole-free, factorizable part of the opposite-helicity Compton amplitudes for any spin.
For QED it is sufficient to simply subtract the terms with unphysical poles from the amplitude, while for QCD and gravity a slightly more involved recursive method---similar to, but simpler than, those of refs.~\cite{Chung:2018kqs,Falkowski:2020aso}---is needed to disentangle physical from unphysical poles, before subtracting the latter.

Curing the factorizable part of the opposite-helicity amplitudes is sufficient for yielding well-defined amplitudes for any spin, but ambiguities in possible contact terms remain to be addressed.
Appealing to our goal of describing Kerr black hole scattering guides us in doing so.
Up to the 2PM order, it has been observed that the spins of black holes are arranged in amplitudes specifically in the combination $q\cdot S_{i}\, q\cdot S_{j}-q^{2}S_{i}\cdot S_{j}$, where $q^{\mu}$ is the momentum transfer and $S_{1,2}^{\mu}$ are the spin vectors of the black holes \cite{Holstein:2008sx,Guevara:2017csg,Damgaard:2019lfh,Bern:2020buy,Kosmopoulos:2021zoq,Chen:2021qkk}.
This correspondence between spin structures---which we will call the {\it black hole spin structure assumption}---has been observed to break upon the inclusion of finite-size effects \cite{Aoude:2020ygw,Kosmopoulos:2021zoq}.
Imposing this assumption on the 2PM amplitude 
is sufficient to uniquely fix the contributions from nearly all potential contact terms that can be added to the opposite-helicity Compton amplitude.
The number of remaining unfixed contact-term coefficients grows modestly with the total spin: we find $\sum_{n=4}^{2s}(\lfloor n/2\rfloor-1)$ such coefficients for a spin $s\geq2$ massive particle (e.g. there are 1, 2, 4 free coefficients for a spin-2, 5/2, 3 particle etc.).
The specific contact terms yielding this correspondence in fact emerge from the recursive approach removing the unphysical poles, and we have confirmed the uniqueness of this set exhaustively up to eighth order in spin.

With this condition for determining contact contributions, we present amplitudes at 2PM up to seventh order in the spin of one object in terms of coefficients parametrizing the few remaining unfixed contact terms.
Complete results up to eighth order in the spin of either object have been relegated to the auxiliary file \texttt{2PMSpin8Aux.nb} due to their cumbersome nature.
We also present results to all orders in spin in the test-mass limit.

As we will emphasize and elaborate below, our results herein benefit greatly from the application of the heavy on-shell variables of ref.~\cite{Aoude:2020onz}.
There are several reasons for this.
First, the separation of scales inherent in these variables makes the classical limit trivial to take.
Second, the difference between the momenta of the initial- and final-state spinors is easily accounted for by these variables, foregoing the need for the so-called Hilbert space matching \cite{Chung:2019duq} and consequently easily manifesting the dependence of the amplitude on the classical spin vector.
From this form of the amplitude, it becomes immediately clear how to isolate the problematic part of the Compton amplitude.
Exhausting the list of possible contact terms at a given spin order and leading order in $\hbar$ is facilitated by the small number of spinor structures when using the heavy on-shell variables.
All-in-all, the benefits of the heavy on-shell variables are neatly illustrated by the fact that the cured Compton amplitudes are notably compact and can be written explicitly, with no implicit recursion remaining.

The structure of this paper is as follows.
We begin in \Cref{sec:OSHPET} with a summary of the heavy on-shell variables of ref.~\cite{Aoude:2020onz} and some of their pertinent properties.
\Cref{sec:HeavyCompton} contains our analysis of the opposite-helicity Compton amplitudes in the classical limit.
The cured amplitudes that we find are presented in \cref{eq:CorrectedQED,eq:CorrectedQCD,eq:CorrectedGR} for QED, QCD, and gravity, respectively.
These amplitudes include the most general contact terms preserving the black hole spin structure assumption mentioned above.
In \Cref{sec:2PM} we make use of the cured gravitational Compton amplitude to compute the 2PM spinning-spinless amplitude exhibiting the black hole spin structure assumption up to seventh order in the spin vector of the spinning object.
We conclude in \Cref{sec:Conclusion}.

\section{On-shell variables for heavy particles}\label{sec:OSHPET}

In ref.~\cite{Aoude:2020onz}, the present authors introduced massive on-shell variables tailored to describing scattering processes involving particles with masses large relative to a characteristic interaction scale.
We summarize some useful facts about these variables here.
Our conventions are given in \Cref{sec:Conventions}.

Consider a particle with four-momentum $p^{\mu}=mv^{\mu}+k^{\mu}$.
Here, the majority of the particle's momentum is carried by the first term, where $v^{\mu}$ is the particle's four-velocity, and the remainder is carried by the residual momentum $k^{\mu}$ satisfying $|k^{\mu}|\ll m$ component-wise.
Such a decomposition of the momentum is applicable to, say, mesons involving one light and one heavy quark relative to the QCD scale---as in heavy quark effective theory (HQET) \cite{Georgi:1990um,HQETRev,Manohar:2000dt}---or to classical gravitational systems---as in heavy black hole effective theory (HBET) \cite{Damgaard:2019lfh}. We will refer to this general class of theories with heavy particles as heavy particle effective theories (HPETs).
In both of these situations, the typical momentum of the light degrees of freedom (gluons and the light quark in the former case, soft gravitons in the latter case) is on the order of the residual momentum.
It was argued in ref.~\cite{Damgaard:2019lfh} that the residual momentum scales with $\hbar$ in the classical limit, based on the $\hbar$ counting scheme of ref.~\cite{Kosower:2018adc}.

Dirac spinors are related to spinors representing heavy spin-1/2 particles through \cite{Damgaard:2019lfh}
\begin{align}
    u_{v}(p)&=\left(1-\frac{\slashed{k}}{2m}\right)u(p),
\end{align}
where the momentum $p^{\mu}$ has been decomposed as above.
Expressing the right-hand side in terms of the massive on-shell variables of ref.~\cite{Arkani-Hamed:2017jhn} and introducing analogous spinors for the heavy states leads to the definition of the heavy on-shell variables \cite{Aoude:2020onz},
\begin{align}\label{eq:OSHPETVars}
    \begin{pmatrix}
	    | p^{I}_{v}\rangle \\
	    | p^{I}_{v} ] 
    \end{pmatrix}
    =
    \left( \mathbb{I} - \frac{\slashed{k}}{2m} \right)
    \begin{pmatrix}
	    | p^{I}\rangle \\
	    | p^{I} ] 
    \end{pmatrix} ,
\end{align}
where $I$ is an $SU(2)$ massive little group index.

Despite describing a state with momentum $p^{\mu}=mv^{\mu}+k^{\mu}$, the heavy on-shell variables correspond to a momentum
\begin{align}\label{eq:HeavyMomentum}
    p_{v}^{\mu}=m_{k}v^{\mu},\quad m_{k}\equiv\left(1-\frac{k^{2}}{4m^{2}}\right)m.
\end{align}
This unintuitive property means that heavy spinors for states whose momenta differ only by a residual momentum (i.e. states of a common velocity) are related by a rescaling.
Specifically, if $p^{\mu}=mv^{\mu}+k^{\mu}$ and $p^{\prime\mu}=mv^{\mu}+k^{\prime\mu}$, then
\begin{align}\label{eq:HeavyVarRescale}
	|p_{v}^{\prime I}\rangle=\sqrt{\frac{m_{k^{\prime}}}{m_{k}}}|p_{v}^{I}\rangle,\quad |p^{\prime}_{vI}]=\sqrt{\frac{m_{k^{\prime}}}{m_{k}}}|p_{vI}].
\end{align}
Note that the little group index is not changed by this relation.
A natural implication of \cref{eq:HeavyMomentum,eq:HeavyVarRescale} is that the heavy on-shell variables should be considered spinors for the four-velocity of the heavy object rather than for its momentum:
\begin{align}\label{eq:VelocitySpinor}
    |p_{v}^{I}\rangle=\sqrt{m_{k}}|v^{I}\rangle,\quad |p_{vI}]=\sqrt{m_{k}}|v_{I}].
\end{align}

Amplitudes expressed in terms of on-shell spinors for heavy particles carry $2s$ symmetrized little group indices for each massive particle in the scattering.
Even for modest values of the spin, writing this symmetrization explicitly results in cumbersome expressions.
It is convenient to suppress this symmetrization by introducing bold notation for massive on-shell variables \cite{Chiodaroli:2021eug}:
\begin{align}\label{eq:BoldNotation}
    |\boldsymbol{v}\rangle\equiv|v^{I}\rangle z_{p,I},\quad |\boldsymbol{v}]\equiv|v^{I}]z_{p,I}.
\end{align}
The $z_{p,I}$ are auxiliary variables that absorb the open little group indices.
In this way, products of bolded spinors are scalar quantities.
Additionally to absorbing little group indices, these variables can be chosen to be the eigenvalues of the spin-coherent states of ref.~\cite{Aoude:2021oqj} under the action of massive little-group creation operators.
With this choice, a rigorous application of the spin-coherent states then has the effect of cancelling all dependence on the auxiliary variables with the normalization of the states \cite{Aoude:2021oqj}.\footnote{This was explicitly shown in ref.~\cite{Aoude:2021oqj} for the three-point amplitude and elastic scattering at tree-level, but we expect it to hold also for the Compton amplitude.}

Defining the spin vector through the Pauli-Lubanski pseudovector with reference momentum $p_{v}^{\mu}$, $S^{\mu}=-\frac{1}{2}\epsilon^{\mu\nu\alpha\beta}v_{\nu}J_{\alpha\beta}$, it's clear that the spin vector is suitable for any heavy state with velocity $v^{\mu}$.
Taking advantage of this, it was shown in ref.~\cite{Aoude:2020onz} that the minimal three-point amplitude in terms of heavy on-shell variables matches immediately onto the descriptions of a Kerr black hole in terms of the one-particle effective action of ref.~\cite{Levi:2015msa} and the "classical amplitude" of ref.~\cite{Vines:2017hyw}.
This was in contrast to other approaches that had to account for the mismatch between the spin vector's reference momentum and the momentum of at least one external matter state \cite{Guevara:2018wpp,Chung:2018kqs,Guevara:2019fsj,Chung:2019duq,Arkani-Hamed:2019ymq,Bern:2020buy}.
Spin contributions to the amplitude are identified through the action of the Pauli-Lubanski pseudovector on irreps of $SL(2,\mathbb{C})$:
\begin{align}\label{eq:Spin1/2Rep}
    {\left(S^{\mu}\right)_{\alpha}}^{\beta}=\frac{1}{4}\left[(\sigma^{\mu})_{\alpha\dot{\alpha}}v^{\dot{\alpha}\beta}-v_{\alpha\dot{\alpha}}(\bar{\sigma}^{\mu})^{\dot{\alpha}\beta}\right],\quad
    {\left(S^{\mu}\right)^{\dot\alpha}}_{\dot\beta}=-\frac{1}{4}\left[(\bar{\sigma}^{\mu})^{\dot{\alpha}\alpha}v_{\alpha\dot\beta}-v^{\dot\alpha\alpha}(\sigma^{\mu})_{\alpha\dot\beta}\right].
\end{align}

The heavy on-shell variables yield four advantages in the context of classical scattering. 
First, through \cref{eq:VelocitySpinor} and the inversion of \cref{eq:OSHPETVars}, the $\hbar\rightarrow0$ limit of the spinors and amplitudes is easy to take.
Second, the compatibility of the spin vector with the heavy variables means that an amplitude expressed in these variables can be easily explicitly recast as a spin-multipole expansion.
Third, the spin vector defined as above can be identified with the spin vector of a classical spinning body, as it automatically satisfies the spin-supplementary condition $p_{v\mu}S^{\mu}=0$ for any state with velocity $v^{\mu}$.
Finally, since initial and final states are represented by spinors for the same velocity, on-shell conditions are immediately applicable, allowing for the simple removal of spinors from classical results.

We can use \cref{eq:VelocitySpinor} to express amplitudes in terms of spinors for the velocity rather than the momenta of the external states.
When doing so, since in the classical limit the initial- and final-state heavy spinors both correspond to external states with velocity $v^{\mu}$, we use $\boldsymbol{\bar{v}}$ to label spinors for the final state.
This keeps track of the differing little-group indices of the initial and final states, which is concealed by the bold notation of \cref{eq:BoldNotation}.

We have written \cref{eq:StandardCompton} using the anti-chiral representation for the massive spinors.
However, when translating to the heavy spinors we can write it in a way that is manifestly independent of the representation.  In particular, the Dirac equation (in the forms $\langle\bar{\boldsymbol{v}}|=-[\bar{\boldsymbol{v}}|v$ and $v|\boldsymbol{v}\rangle=|\boldsymbol{v}]$) in combination with the on-shell conditions in \cref{eq:HQETOnShellness} for heavy spinors gives
\begin{align}\label{eq:SpinorProduct}
    \langle\bar{\boldsymbol{v}}\boldsymbol{v}\rangle=\bar{z}_{v,I}z_{v}^{J}\langle v^{I}v_{J}\rangle=-\bar{z}_{v,I}z_{v}^{J}[v^{I}v_{J}]=\bar{z}_{v,I}z_{v}^{I}\equiv |z_{v}|^{2}.
\end{align}
Also, from \cref{eq:Spin1/2Rep}, we can see that
\begin{align}
    {\left(S^{\mu}\right)_{\alpha}}^{\beta}&=v_{\alpha\dot{\alpha}}{\left(S^{\mu}\right)^{\dot\alpha}}_{\dot\beta}v^{\dot{\beta}\beta},\quad {\left(S^{\mu}\right)^{\dot\alpha}}_{\dot\beta}=v^{\dot{\alpha}\alpha}{\left(S^{\mu}\right)_{\alpha}}^{\beta}v_{\beta\dot{\beta}}.
\end{align}
Introducing new bracket notation,\footnote{This is reminiscent of the generalized expectation value of refs.~\cite{Guevara:2018wpp,Guevara:2019fsj}.}
\begin{align}
	\langle T^{\mu} \rangle\equiv\frac{\langle\bar{\boldsymbol{v}}|T^{\mu} |\boldsymbol{v}\rangle}{\langle\bar{\boldsymbol{v}}\boldsymbol{v}\rangle},\quad [T^{\mu}]\equiv\frac{[\bar{\boldsymbol{v}}|T^{\mu}|\boldsymbol{v}]}{[\bar{\boldsymbol{v}}\boldsymbol{v}]},
\end{align}
where $T$ is any matrix transforming in the $(1/2,1/2)$ representation of $SL(2,\mathbb{C})$, this gives $\langle S^{\mu}\rangle=[S^{\mu}]$.
The identical relation holds for the ring radius, $a^{\mu} \equiv S^{\mu}/m$, giving $\mathfrak{a}^{\mu}\equiv\langle a^{\mu}\rangle=[a^{\mu}]$.
Representation independence is manifested by writing the amplitude in terms of $\mathfrak{a}^{\mu}$.

\Cref{eq:SpinorProduct} implies that an amplitude expressed in terms of the heavy on-shell variables will come with an overall factor of $|z_{v}|^{4s}$.
Since, as mentioned above, the spin-coherent states of ref.~\cite{Aoude:2021oqj} will cancel this overall factor, we simply omit it from the amplitudes in the rest of the paper.

Before moving on to the Compton amplitude, let us summarize our rules for counting factors of $\hbar$.
Momenta corresponding to massless fields scale with one factor of $\hbar$ in the classical limit \cite{Kosower:2018adc}.
As a consequence of this, the on-shell spinors for massless fields scale as $\sqrt{\hbar}$.
To preserve the dimensions of the amplitude when $\hbar$ is restored, the spin vector must scale as $\hbar^{-1}$ \cite{Maybee:2019jus}.
We will see below that each power of the spin vector comes with one power of a momentum scaling with $\hbar$, so that, in total, the spin terms appearing in the Compton amplitude do not scale with $\hbar$.
These rules mean that $\{t_{1i},y\}\sim\hbar$ and $s_{34}\sim\hbar^{2}$.
Consequently, the QED and gravitational Compton amplitudes are $\mathcal{O}(\hbar^{0})$ at leading order, and the QCD partial amplitudes are $\mathcal{O}(\hbar^{-1})$ at leading order (we ignore $\hbar$ factors from coupling constants).

From this point forth, unless explicitly stated otherwise, we will use "Compton amplitude" to refer specifically to the opposite-helicity Compton amplitude.

\section{The Compton amplitude}\label{sec:HeavyCompton}

In this section, we present our analysis of \cref{eq:StandardCompton} and the resulting spurious-pole-free classical Compton amplitudes.
To begin, let us write the amplitudes of interest in terms of the heavy on-shell variables in \cref{eq:OSHPETVars}.

Abbreviating the momentum arguments and writing the amplitudes with a unified notation, \cref{eq:StandardCompton} becomes\footnote{The exponential is to be understood in terms of its Taylor series, which truncates after $2s+1$ terms for finite spin $s$; for details, see e.g. ref.~\cite{Guevara:2018wpp}. This structure arises when the spin vector is written in the spin-$s$ representation, as opposed to the spin-$1/2$ representation in \cref{eq:Spin1/2Rep}. We write the ring radius $\mathfrak{a}^{\mu}$ instead of $S^{\mu}$ when working in the spin-$s$ representation.}
\begin{align}\label{eq:HeavyCompton}
    \cA^{s}&=\left(\frac{m}{m_{q}}\right)^{s}\cA^{0}\exp\left\{\left[q_{4}-q_{3}+\frac{(t_{14}-t_{13})}{y}w\right]\cdot\mathfrak{a}\right\},
\end{align}
where $q^{\mu}\equiv q_{3}^{\mu}+q_{4}^{\mu}$ and $w^{\mu}\equiv[4|\bar \sigma^{\mu}|3\rangle/2$.
The spin-0 gauge theory or gravity amplitude $\cA^{0}$ is obtained by setting $s=0$ in the appropriate amplitude in \cref{eq:StandardCompton}.
We have used \cref{eq:VelocitySpinor} to express the amplitude in terms of spinors for the velocity rather than the momenta of the external states.
The heavy spinors have all been absorbed into the definition of $\mathfrak{a}^{\mu}$ in \Cref{sec:OSHPET}, manifesting the basis-independence of the amplitude.

A truncation in $\hbar$ has been performed in \cref{eq:HeavyCompton} in the conversion from the spin-1/2 to the spin-$s$ representation of the spin vector.\footnote{We are grateful to Lucile Cangemi for pointing this out.}
The completion of the classical limit is easily obtained by replacing $m_{q}\rightarrow m$.
This form of the amplitude presents two advantages over the equivalent expression in \cref{eq:StandardCompton}.
First, as advertised, expressions in the heavy on-shell variables have facilitated an explicit spin-multipole expansion of the amplitude.
Second, all occurrences of the spurious pole are brought about by the third term in the exponential, isolating the problematic part of the amplitude.
However, we point out that employing the heavy on-shell variables has restricted us to the regime $|q^{\mu}_{3,4}|\ll m$.

Expanding the exponential in \cref{eq:HeavyCompton} in powers of $\mathfrak{a}^{\mu}$, spurious poles in $y$ contaminate the amplitude starting at cubic order in spin for gauge theory and quintic order for gravity.
In the following, we demonstrate the removal of these spurious poles for QED, QCD, and gravity in the classical limit and for any spin, without affecting the factorization properties of the amplitudes on physical poles.
The cured amplitudes that we obtain will contain the most general set of contact terms that obey the black hole spin structure assumption when used to construct the 2PM amplitude.

\subsection{QED}\label{sec:QEDCompton}

We begin with fixing the higher-spin QED Compton amplitudes in the classical limit.
The exponential in \cref{eq:HeavyCompton} can be separated such that the portion without unphysical poles is simply an overall factor.
In the classical limit,\footnote{The exponential can be decomposed without use of the Baker-Campbell-Hausdorff formula because the exponent is just a number.}
\begin{align}
	\label{eq:QEDclassical}
	\lim_{\hbar\rightarrow0}\cA^{s}_{\text{QED}}\equiv \cA_{\text{cl},\text{QED}}^{s}&=e^{(q_{4}-q_{3})\cdot\mathfrak{a}}\sum_{n=0}^{2s}\frac{1}{n!}I_{n},
\end{align}
where
\begin{align}
    I_{n}&\equiv\frac{1}{t_{13}t_{14}}\frac{(t_{14}-t_{13})^{n}}{y^{n-2}}(w\cdot\mathfrak{a})^{n}.
\end{align}
We must keep in mind that there cannot be more than $2s$ powers of $\mathfrak{a}^{\mu}$ in any term in the amplitude for a massive spin-$s$ particle.
For $n>2$, the sum produces unphysical poles in $y$.
Such a term can be written as
\begin{align}\label{eq:QEDAmpFixedn}
    I_{n\geq2}=-4\frac{(t_{14}-t_{13})^{n-2}}{y^{n-2}}\left(w\cdot\mathfrak{a}\right)^{n}+\mathcal{O}(\hbar).
\end{align}
We have used that the Mandelstam variables are related via $t_{13}=-t_{14}+\mathcal{O}(\hbar^{2})$.
Consequently, none of the terms with unphysical poles contribute to the physical factorization channels at leading order in $\hbar$.
This allows us to write a spurious-pole-free amplitude simply by subtracting off the residues on unphysical poles, amounting to dropping all terms in the sum in \cref{eq:QEDclassical} with $n>2$.

One can freely add contact terms to the spurious-pole-free amplitude without modifying the factorization properties of the amplitude.
Therefore, simply removing the spurious poles for $n>2$ does not result in a unique QED Compton amplitude.
Let us now elaborate on the contact terms that can deform the Compton amplitude at leading order in $\hbar$.

The form of potential contact terms is very constrained in the classical limit by their little group and $\hbar$ scalings.
First, the little group weights for the photons must be carried by either $y$ or $\wa$.
Then, the $\hbar$ scaling of the amplitude tells us how these structures should be combined with other spin structures and Mandelstams.
The overall $\hbar$ scaling of the amplitude is $\hbar^0$ (when ignoring the $\hbar$'s from the couplings), while $y$ scales as $\hbar$ and $\wa$ scales as $\hbar^{0}$.
Since there can be no compensating denominators for the contact terms, $y$ must be grouped with factors of the ring radius $\mathfrak{a}$, which scale as $1/\hbar$.
Therefore, the little group weight must come from one of the three factors\footnote{We neglect contact terms with fewer powers of $\hbar$ than the factorizable part of the amplitude.}
\begin{align}
	\label{eq:ContactLG}
	\frac{y^2}{m^{2}} \mathfrak{a}^{2} , \qquad 
	y (\wa) \frac{(t_{14} - t_{13})}{m^{2}} \mathfrak{a}^{2} , \qquad 
	(\wa)^2 .
\end{align}
In the second factor we multiplied by a difference of Mandelstams such that the whole factor scales as $\hbar^{0}$.
Since $t_{13}+t_{14}=\mathcal{O}(\hbar^{2})$, the above combination of Mandelstams is the only one at $\mathcal{O}(\hbar)$.
The masses give these factors the same dimensions as the amplitude.

Little group constraints bar any contact-term deformations of the Compton amplitude at leading order in $\hbar$ at zeroth or linear order in spin.
Higher-spin contact terms can be obtained by dressing the factors in \cref{eq:ContactLG} with dimensionless spin structures that scale as $\hbar^{0}$.
The possible dressing monomials are
\begin{align}
	\label{eq:ContactMonomial}
	q_3\cdot \mathfrak{a}, \qquad 
	q_4\cdot \mathfrak{a}, \qquad 
	s_{34} \mathfrak{a}^{2}, \qquad 
	\frac{(t_{14} - t_{13})^2}{m^{2}} \mathfrak{a}^{2}, \qquad 
\end{align}
A general contact term at a fixed order in spin is given by multiplying one of the factors in \cref{eq:ContactLG} by monomials from \cref{eq:ContactMonomial} until the desired spin order is reached.
The full set of contact terms (including redundancies) is obtained by forming all possible such combinations.
Assigning each contact term its own coefficient leads to a rapid proliferation of independent coefficients as the spin increases.

This vast space of contact terms can be reduced substantially by imposing that the black hole spin structure assumption described in \Cref{sec:Introduction} persists in the 2PM amplitude describing classical scattering at all spins.
As a reminder to the reader, all calculations of black hole scattering up to third order in the spin vectors\footnote{The results of ref.~\cite{Chen:2021qkk} at fourth order in spin also exhibit this correspondence. We don't include them in this breath, however, since contact terms can be introduced that spoil the correspondence without introducing new scales to the theory.} have shown a specific spin structure, in that the result only depends on \cite{Holstein:2008sx,Guevara:2017csg,Damgaard:2019lfh,Bern:2020buy,Kosmopoulos:2021zoq,Chen:2021qkk}
%
\begin{align}
	\label{eq:spinStructure}
	(q\cdot \mathfrak{a}_{i}) (q\cdot \mathfrak{a}_{j}) - q^2 (\mathfrak{a}_{i}\cdot \mathfrak{a}_{j})  , \qquad i,j = 1,2,
\end{align}
and not on these two structures independently.
Here $q$ is the transferred momentum and the spins can be those of either spinning body.
This is in contrast to including higher-curvature operators in the scattering, which have the interpretation of describing finite-size effects and which spoil the correspondence between the spin structures \cite{Aoude:2020ygw,Kosmopoulos:2021zoq}.
Analogously, non-minimal couplings in the three-point amplitude require the introduction of new (length) scales if they are to contribute finitely in the classical limit \cite{Aoude:2021oqj}.

This observation guides us in our construction of the Compton amplitude.
Allowing for the possibility that this is not an accidental phenomenon for low spin orders, we investigate the implications for the Compton amplitude if we fix the coefficients of contact terms such that this property is obeyed.
As we will explain in more detail in \Cref{app:spinStructure}, the only groupings of the spins in the Compton amplitude which respect \cref{eq:spinStructure} at 2PM are $\wa$ and
\begin{subequations}\label{eq:s1s2}
\begin{align}
	\mathfrak{s}_1 &\equiv (q_3 - q_4 )\cdot \mathfrak{a} , \\
	\mathfrak{s}_{2} &\equiv -4 (q_3 \cdot \mathfrak{a})(q_4 \cdot \mathfrak{a}) + s_{34} \mathfrak{a}^{2} .
\end{align}
\end{subequations}
Contact terms must therefore arise in proportions that can be written as products of these three structures.
We observe that most contact-term coefficients are uniquely fixed by this requirement.

With these considerations, the spurious-pole-free QED Compton amplitude is
\begin{subequations}\label{eq:CorrectedQED}
\begin{align}
	\cA_{\text{cl},\text{QED}}^{s}&=e^{(q_{4}-q_{3})\cdot\mathfrak{a}}\sum_{n=0}^{2s}\frac{1}{n!}\bar I_{n},
\end{align}
where
\begin{align}
	\bar I_{n} \equiv
	\begin{cases}
	    I_{n}, & n<2 \\
    	I_{2} + (\wa)^2c_{0}^{(2)} , &  n= 2, \\
    	(\wa)^2 \sum_{j=0}^{\floor{(n-2)/2}} c_{j}^{(n)} \mathfrak{s}_{1}^{n-2-2j} \mathfrak{s}_{2}^{j} , &  n> 2 .
	\end{cases}
\end{align}
\end{subequations}
For a given $n$ there are $\floor{n/2}$ free coefficients $c_{j}^{(n)}$.
\Cref{eq:CorrectedQED} contains the classical limit of the spin-3/2 gauge theory result of ref.~\cite{Chiodaroli:2021eug}.
The amplitude there is obtained by truncating the sum in $n$ in \cref{eq:CorrectedQED} at $n=3$ and setting $c_{0}^{(2)}=4/3$ and $c_{0}^{(3)}=0$.
\Cref{eq:StandardCompton,eq:HeavyCompton} have $c_{0}^{(2)}=0$ at quadratic-order in spin .

\subsection{QCD}\label{sec:QCDCompton}

As a step between the electromagnetic and gravitational amplitudes, we consider now the QCD case.
Similarly to gravity, the QCD amplitude possesses a factorization channel on a massless pole.
While a spurious-pole-free QED Compton amplitude for all spins could be obtained immediately, the presence of the massless pole necessitates a different, recursive approach.

This method is based on the relation
\begin{align}
	\label{eq:HeavyRecursion}
	(t_{14} - t_{13})^{2} (\wa)^2 = - 4m^2 s_{34} (\wa)^2 + 2 y (t_{14} - t_{13}) \mathfrak{s}_{1} (\wa) - y^2 \mathfrak{s}_{2} ,
\end{align}
which is a manifestation of the vanishing of the Gram determinant in four spacetime dimensions for the five four-vectors $w^{\mu},\,q_{3}^{\mu},\,q_{4}^{\mu},\,p_{1}^{\mu},$ and $\mathfrak{a}^{\mu}$.
In the context of the Compton amplitude, \cref{eq:HeavyRecursion} allows us to trade terms singular in both $s_{34}$ and $y$ for terms which either have no poles in physical channels or have lower-order poles in $y$.
Therefore, applying \cref{eq:HeavyRecursion} recursively results in the elimination of all terms with both spurious poles and non-vanishing residues on physical poles.
Then, a cured all-spin amplitude can be obtained by dropping all terms with spurious poles.

Let us be explicit about this.
Consider the QCD partial amplitudes in the classical limit, which we denote without calligraphic font:
\begin{align}
	A_{\text{cl},\text{QCD}}(-\mathbf{1}^{s}_{i},\mathbf{2}^{s}_{j},3^{-}_{c},4^{+}_{d})&=e^{(q_{4}-q_{3})\cdot\mathfrak{a}}\sum_{n=0}^{2s}\frac{1}{n!}J_{n}, \\
	A_{\text{cl},\text{QCD}}(-\mathbf{1}^{s}_{i},\mathbf{2}^{s}_{j},4^{+}_{d},3^{-}_{c})&=-e^{(q_{4}-q_{3})\cdot\mathfrak{a}}\sum_{n=0}^{2s}\frac{1}{n!}J_{n},
\end{align}
where
\begin{align}
    J_{n}\equiv\frac{y^{2}}{s_{34}t_{14}}\left(\frac{t_{14}-t_{13}}{y}w\cdot\mathfrak{a}\right)^{n}.
\end{align}
In the classical limit, one power of $t_{14}-t_{13}$ can be cancelled with the pole in $t_{14}$, up to an overall factor.
Thus, \cref{eq:HeavyRecursion} implies the existence of a recursion relation for $J_{n}$ for $n\geq3$:
\begin{align}\label{eq:QCDRecursion}
    J_{n}&=-8m^{2}\frac{(t_{14}-t_{13})^{n-3}}{y^{n-2}}(w\cdot\mathfrak{a})^{n}+2\mathfrak{s}_{1}J_{n-1}-\mathfrak{s}_{2}J_{n-2}. 
\end{align}
This relation can be solved exactly for $J_{n}$.
The result is, however, quite cumbersome, and contains the spurious poles we wish to eliminate.
We can simplify the solution to the recursion by focusing first on the factorizable part of the amplitude.
Modifying the recursion to
\begin{align}\label{eq:FactorizableRecursionQCD}
    J_{n\geq3}&\sim-2[(q_{4}-q_{3})\cdot\mathfrak{a}]J_{n-1}+4(q_{3}\cdot\mathfrak{a})(q_{4}\cdot\mathfrak{a})J_{n-2},
\end{align}
the solution will not contain any spurious poles at the cost of also dropping contact terms.
We will restore the latter later such that the black hole spin structure assumption is respected.

We can now solve \cref{eq:FactorizableRecursionQCD} for the factorizable part of the all-spin QCD Compton amplitude.
The solution is
%
\begin{align}\label{eq:RecursionSolution}
    J_{n\geq3}&\rightarrow
    J_{1}2^{n-1}(q_{3}\cdot\mathfrak{a})(q_{4}\cdot\mathfrak{a})\sum_{j=0}^{n-3}(q_{3}\cdot\mathfrak{a})^{j}(-q_{4}\cdot\mathfrak{a})^{n-3-j}+J_{2}2^{n-2}\sum_{j=0}^{n-2}(q_{3}\cdot\mathfrak{a})^{j}(-q_{4}\cdot\mathfrak{a})^{n-2-j}.
\end{align}
When we discarded contact terms, we decomposed the spin structures from $\mathfrak{s}_{1}$ and $\mathfrak{s}_{2}$ to polynomials in $q_3\cdot \mathfrak{a}$ and $q_4\cdot \mathfrak{a}$, thus violating the black hole spin structure assumption at one-loop level.
We can remedy this by adding the appropriate contact terms, amounting to the replacements
\begin{subequations}\label{eq:Lrecursion}
\begin{align}
	(q_3 \cdot \mathfrak{a}) (q_4\cdot \mathfrak{a}) &\rightarrow -\frac{1}{4} \mathfrak{s}_{2} , \\
	\sum_{j=0}^{m} (q_3 \cdot \mathfrak{a})^{j} (-q_4\cdot \mathfrak{a})^{m-j} &\rightarrow  2^{-m} L_{m} ,
	\label{eq:restoreS1S2}
\end{align}
where
\begin{align}
	L_{m} \equiv \sum_{j=0}^{\floor{m/2}} \binom{m+1}{2j+1} \mathfrak{s}^{m-2j}_{1} (\mathfrak{s}^2_1 - \mathfrak{s}_{2})^{j} .
\end{align}
\end{subequations}
With these restorations, we find the {\it unique} QCD Compton amplitude producing one-loop amplitudes satisfying the black hole spin structure assumption:
\begin{subequations}\label{eq:CorrectedQCD}
\begin{align}
	A_{\text{cl},\text{QCD}}(-\mathbf{1}^{s}_{i},\mathbf{2}^{s}_{j},3^{-}_{c},4^{+}_{d})&=e^{(q_{4}-q_{3})\cdot\mathfrak{a}}\sum_{n=0}^{2s}\frac{1}{n!}\bar{J}_{n}, \\
	A_{\text{cl},\text{QCD}}(-\mathbf{1}^{s}_{i},\mathbf{2}^{s}_{j},4^{+}_{d},3^{-}_{c})&=-e^{(q_{4}-q_{3})\cdot\mathfrak{a}}\sum_{n=0}^{2s}\frac{1}{n!}\bar{J}_{n},
\end{align}
where
\begin{align}
	\bar J_{n} \equiv
	\begin{cases}
	    J_{n} , & n\leq 2 \\
	    J_{2} L_{n-2} - J_{1} \mathfrak{s}_{2} L_{n-3} , &  n > 2 .
	\end{cases}
\end{align}
\end{subequations}

Uniqueness is a consequence of the amplitude scaling as $\hbar^{-1}$.
This implies that any contact term at leading order in $\hbar$ must be proportional to some positive power of $\mathfrak{a}^{2}$, but such contact terms cannot be added freely without violating the black hole spin structure assumption.
There is thus no freedom to add contact terms when the black hole spin structure assumption is imposed, in contrast to QED, and \cref{eq:CorrectedQCD} is the unique classical limit of the spurious-pole-free QCD Compton amplitude for all spins under this condition.

\subsection{Gravity}\label{sec:GRCompton}

Finally, let us consider the gravitational case.
The situation now remains very similar to that of QCD: the presence of the massless factorization channel necessitates the application of \cref{eq:HeavyRecursion}.
We write the arbitrary-spin amplitude as
\begin{align}
	\cM_{\text{cl}}^{s}&=e^{(q_{4}-q_{3})\cdot\mathfrak{a}}\sum_{n=0}^{2s}\frac{1}{n!}K_{n},
\end{align}
where
\begin{align}
    K_{n}&\equiv\frac{y^{4}}{s_{34}t_{13}t_{14}}\left(\frac{t_{14}-t_{13}}{y}w\cdot\mathfrak{a}\right)^{n}.
\end{align}
Then, through \cref{eq:HeavyRecursion}, this obeys the recursion relation
\begin{align}\label{eq:GRRecursion}
    K_{n}&=16m^{2}\frac{(t_{14}-t_{13})^{n-4}}{y^{n-4}}(w\cdot\mathfrak{a})^{n}+2\mathfrak{s}_{1}K_{n-1}-\mathfrak{s}_{2}K_{n-2}. 
\end{align}
This is the same recursion relation as \cref{eq:QCDRecursion}, but now it is applied for $n>4$.
The solution to this recursion for $n>4$ is, after ignoring contact terms and terms with only spurious poles,
\begin{align}\label{eq:RecursionSolutionGR}
    K_{n}&\rightarrow 2^{n-2}(q_{3}\cdot\mathfrak{a})(q_{4}\cdot\mathfrak{a})K_{2}\sum_{j=0}^{n-4}(q_{3}\cdot\mathfrak{a})^{j}(-q_{4}\cdot\mathfrak{a})^{n-4-j}+2^{n-3}K_{3}\sum_{j=0}^{n-3}(q_{3}\cdot\mathfrak{a})^{j}(-q_{4}\cdot\mathfrak{a})^{n-3-j}.
\end{align}
We can ensure that the black hole spin structure assumption is satisfied in the same way as for QCD by performing the replacements in \cref{eq:Lrecursion}. 
In addition, further contact terms can be added in analogous forms as for QED since the amplitude scales as $\hbar^0$.

The most general gravitational amplitude free of unphysical poles and obeying the black hole spin structure assumption at 2PM is
\begin{subequations}\label{eq:CorrectedGR}
\begin{align}
	\cM_{\text{cl}}^{s}&=e^{(q_{4}-q_{3})\cdot\mathfrak{a}}\sum_{n=0}^{2s}\frac{1}{n!}\bar{K}_{n},
\end{align}
where
\begin{align}
    \bar{K}_{n}\equiv
    \begin{cases}
        K_{n}, & n<4, \\
        K_{4} + m^{2}(\wa)^4 d_{0}^{(4)} , & n=4, \\
        K_{3} L_{n-3}  - K_{2} \mathfrak{s}_{2} L_{n-4} + m^{2}(\wa)^4 \sum_{j=0}^{\floor{(n-4)/2}} d_{j}^{(n)} \mathfrak{s}_{1}^{n-4-2j} \mathfrak{s}_{2}^{j}, & n>4.
    \end{cases}
\end{align}
\end{subequations}
For a given $n$ there are $\floor{(n-2)/2}$ free coefficients $d_{j}^{(n)}$.
The part of the $\bar{K}_{n}$ with no free coefficients is in fact the solution to \cref{eq:GRRecursion} once the first term on the right-hand side (carrying all spurious poles) is dropped.
\Cref{eq:CorrectedGR} also clearly exhibits spin-multipole universality.
Each $\bar{K}_{n}$ is $\mathcal{O}(\mathfrak{a}^{n})$, and contributes only at this spin order.
Therefore, increasing the spin only adds new terms to the sum without affecting lower spin multipoles.

\Cref{eq:CorrectedGR} contains the classical limit of the spin-5/2 gravitational Compton amplitude of refs.~\cite{Falkowski:2020aso,Chiodaroli:2021eug}.
It is obtained by setting $d_{0}^{(4)}=-16/5$ and $d_{0}^{(5)}=32$ and truncating the sum at $n=5$.
Notably, as the Compton amplitude presented there was the best behaved in the high-energy limit, this means that the best behaved amplitude exhibits \cref{eq:spinStructure} when used to construct the 2PM amplitude.
\Cref{eq:StandardCompton,eq:HeavyCompton} have $d_{0}^{(4)}=0$; we will come back to this difference in the next section.

Little group scaling bars any contact-term deformations of the leading-in-$\hbar$ Compton amplitude at spin orders $\mathfrak{a}^{n\leq3}$, unless they are accompanied by dimensionful coefficients, as in refs.~\cite{Haddad:2020que,Aoude:2020ygw}.

The infinite-spin limits for the Compton amplitudes in \cref{eq:CorrectedQED,eq:CorrectedQCD,eq:CorrectedGR} are trivial to take, simply extending the finite sum over $n$ to an infinite sum.

\section{2PM scattering to seventh order in spin}\label{sec:2PM}

The Compton amplitude in \cref{eq:CorrectedGR} allows us to compute the 2PM amplitude describing the gravitational scattering of two spinning objects to any order in the spins of the objects.
In the body of this paper we present results up to seventh order in the spin of one object.
The auxiliary file \texttt{2PMSpin8Aux.nb} contains full results up to eighth order in the spin of either scattering object.
This extends previous work on this amplitude, which has been limited to fourth order in the spin of either object \cite{Guevara:2017csg,Guevara:2018wpp,Chung:2018kqs,Damgaard:2019lfh,Bern:2020buy,Kosmopoulos:2021zoq,Chen:2021qkk}.

\begin{figure}
\centering
\subfloat{
    \includegraphics[scale=0.4]{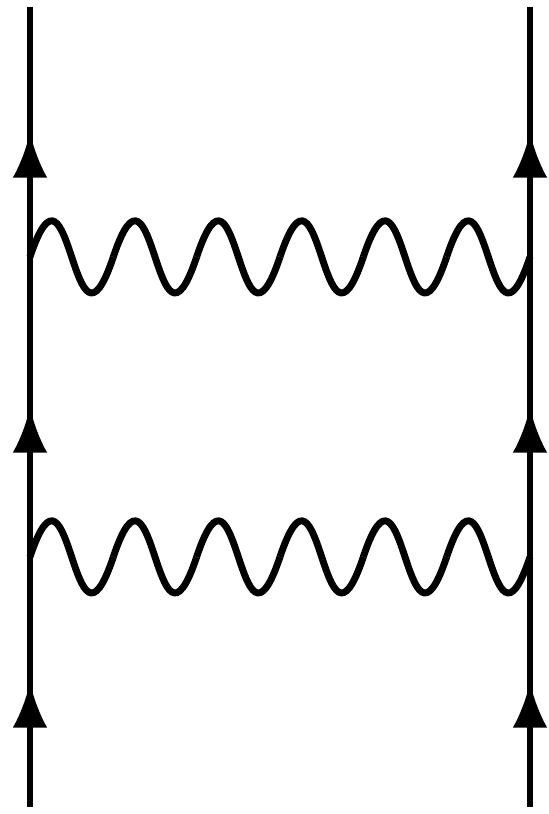}
    }
\hspace{1cm}
\subfloat{
    \includegraphics[scale=0.4]{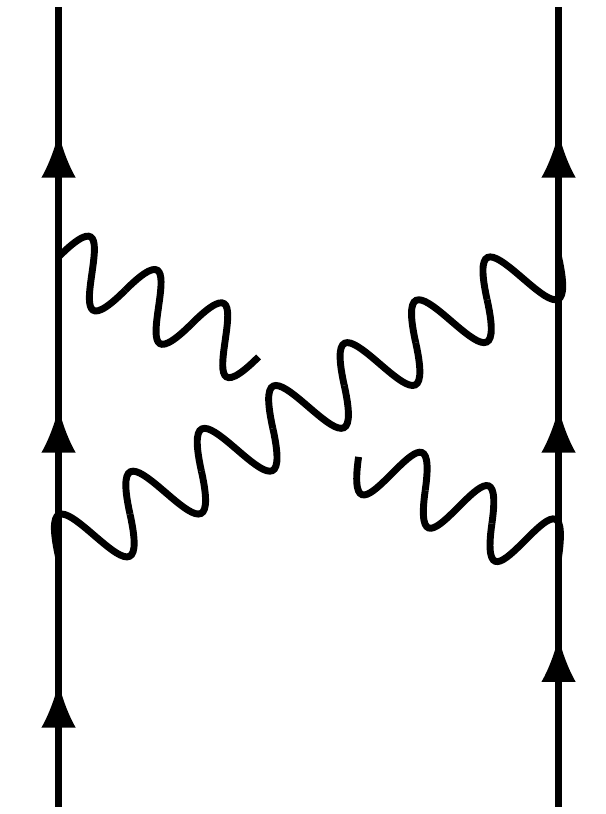}
    }
\hspace{1cm}
\subfloat{
    \includegraphics[scale=0.4]{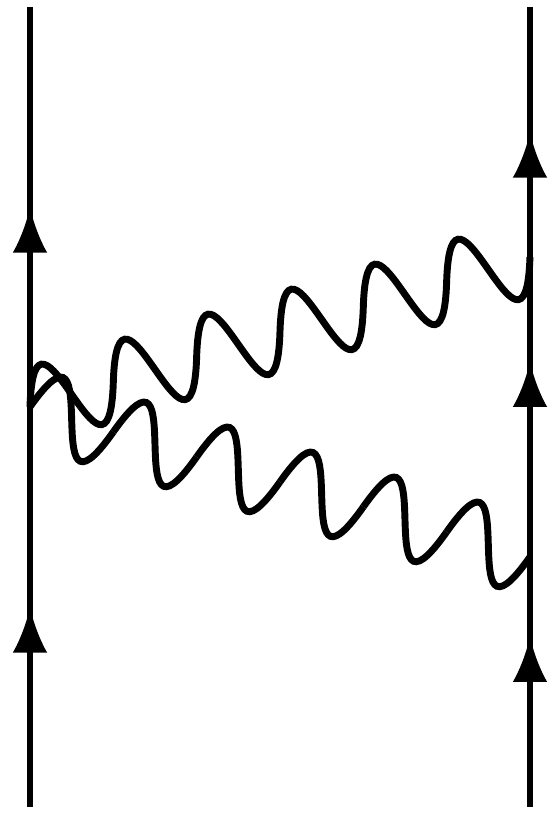}
    }
\hspace{1cm}
\subfloat{
    \includegraphics[scale=0.4]{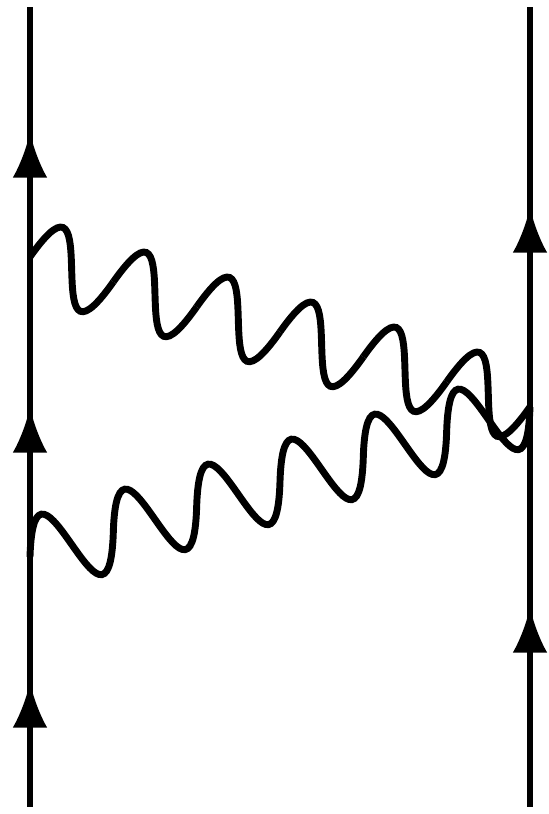}
    }
\caption{\label{fig:OneLoopTopologies}The topologies forming the basis for the 2PM amplitude.}
\end{figure}

Starting with the Compton amplitude, the most economic way to construct the 2PM amplitude is through generalized unitarity \cite{Bern:1994zx,Bern:1997sc,Bern:1994cg,Bern:1998ug,Britto:2004nc,Bern:2007ct}. 
As we are interested in describing long-range interactions, we must evaluate the part of the one-loop amplitude with two graviton lines in the loop \cite{Donoghue:1993eb,Holstein:2004dn}.
Moreover, restricting our focus to classical effects, we need only consider topologies with at least one matter line per loop \cite{Neill:2013wsa}.
Hence, the classical part of the one-loop amplitude can be decomposed in a basis of scalar box, cross-box, and triangle integrals (see \cref{fig:OneLoopTopologies}):
\begin{align}
    \mathcal{M}_{2\text{PM}} = c_\Box {\cal I}_\Box + c_{\crossBox} {\cal I}_{\crossBox} +  c_\triangleleft {\cal I}_\triangleleft + c_\triangleright {\cal I}_\triangleright.
\end{align}
The subscript reminds us that we are at $\mathcal{O}(G^{2})$.
In fact, the classical contributions to the amplitude will come only from the triangle coefficients, whereas the box and cross-box contributions will cancel with analogous contributions from the EFT computing the interaction potential \cite{Cheung:2018wkq,Bern:2020buy} (alternatively, they will cancel with the Born subtraction \cite{Holstein:2008sx,Cristofoli:2019neg}).
As the scalar box and cross-box integrals do not produce classical effects, we will focus solely on the triangle coefficients.
The corresponding integrals---performed in the soft region where the loop momentum satisfies $|l|\ll m_{1,2}$ for the mass of either particle---are
\begin{subequations}
\begin{align}
    \mathcal{I}_{\triangleleft}&=\int\frac{d^{4}l}{(2\pi)^{4}}\frac{1}{l^{2}(q+l)^{2}[(p_{1}-l)^{2}-m_{1}^{2}]}=-\frac{i}{32m_{1}\sqrt{-q^{2}}}, \\
    \mathcal{I}_{\triangleright}&=\int\frac{d^{4}l}{(2\pi)^{4}}\frac{1}{l^{2}(q+l)^{2}[(p_{2}+l)^{2}-m_{2}^{2}]}=-\frac{i}{32m_{2}\sqrt{-q^{2}}}.
\end{align}
\end{subequations}
Here, $q^{\mu}$ is the transfer momentum and our kinematics are
\begin{align}
    p_1^{\mu} = m_1 v_1^{\mu},\qquad p_2^{\mu} = m_2 v_2^{\mu}, \qquad p_1^{\prime\mu} = m_1 v_1^{\mu} + q^{\mu}, \qquad p_2^{\prime\mu} = m_2 v_2^{\mu} - q^{\mu}.
\end{align}
The $v_{i}^{\mu}$ satisfy $v_{i}^{2}=1$, implying the on-shell conditions
\begin{align}
    v_{1}\cdot q=-\frac{q^{2}}{2m_{1}},\qquad v_{2}\cdot q=\frac{q^{2}}{2m_{2}}.
\end{align}

Our task is now to determine the coefficients $c_{\triangleleft,\triangleright}$.
We apply the formalism of refs.~\cite{Forde:2007mi,Kilgore:2007qr} to do so.
The starting point is a $t$-channel cut of both gravitons in the loop, expressed as a product of Compton amplitudes:
\begin{align}
\delta_2 \mathcal{M}_{2\text{PM}}^{(s_1,s_2)} = \sum_{h_1,h_2=\pm 2} \mathcal{M}(-{\bf 1}^{s_1},{\bf 1}^{\prime s_1},l^{h_1},-(l+q)^{h_2})
\mathcal{M}(-{\bf 2}^{s_2},{\bf 2}^{\prime s_2},-l^{-h_1},(l+q)^{-h_2}).
\end{align}
The triangle coefficients that we are after are isolated by cutting one internal massive propagator as well.
This can be obtained from the two-line cut above through
\begin{align}
    C_{3,\triangleleft}^{(s_{1},s_{2})} =-(2p_1\cdot l)\,\delta_2 \mathcal{M}_{2\text{PM}}^{(s_{1},s_{2})}\rvert_{p_1\cdot l \rightarrow 0},\qquad C_{3,\triangleright}^{(s_{1},s_{2})} =(2p_2\cdot l)\,\delta_2 \mathcal{M}_{2\text{PM}}^{(s_{1},s_{2})}\rvert_{p_2\cdot l \rightarrow 0}.
\end{align}
The latter can in fact be obtained from the former through the substitutions $p_{1}\leftrightarrow p_{2},\, m_{1}\leftrightarrow m_{2},\, \mathfrak{a}_{1}\leftrightarrow \mathfrak{a}_{2},\, q\rightarrow-q$.
We will phrase the rest of the discussion in terms of the former cut.

Cutting internal lines means placing their momenta on-shell, imposing three conditions on the loop momentum:
\begin{align}
	\label{eq:trianglecut}
    l^{2}=0,\qquad q\cdot l=-\frac{q^{2}}{2},\qquad p_{1}\cdot l=0.
\end{align}
In four spacetime dimensions, this leaves one free parameter, $t$, in the two solutions for the loop momentum that satisfy the three cut conditions; see e.g. refs.~\cite{Bern:2020buy,Chen:2021qkk} for explicit forms of the loop momentum solutions.
The triangle coefficients are finally obtained by averaging over the triple cuts with both loop momentum solutions, and isolating the part that is independent of the free parameter:
\begin{align}
    c_{\triangleleft}^{(s_{1},s_{2})}&=\frac{1}{2}\sum_{l=l_{1,2}}\left.C_{\triangleleft}^{(s_{1},s_{2})}(l)\right|_{t^{0}},
\end{align}
where the subscript $t^{0}$ represents the constant term of the Laurent series around $t=\infty$ or $t=0$.

For all results in the proceeding subsections, we have computed the 2PM amplitude in this way using two sets of contact terms in the Compton amplitude. 
First, we used the Compton amplitude in \cref{eq:CorrectedGR}.
We confirmed that, as argued in \Cref{sec:GRCompton}, this Compton amplitude always produces contributions to the 2PM amplitude that obey the black hole spin structure assumption.
Then, we computed by including the most general set of contact terms at leading order in $\hbar$ in the Compton amplitude,\footnote{The most general sets we used were not necessarily free of redundancies, but this has no bearing on the statements in this section.} and fixed the coefficients of these contact terms by imposing the black hole spin structure assumption.
Comparing both calculations allowed us to exhaustively check whether the set of contact terms satisfying this property is uniquely that present in \cref{eq:CorrectedGR}.

It has recently been demonstrated that the same-helicity Compton amplitude can contribute to the triangle cut when deviations from Kerr at three-points---that is, deviations from the minimal amplitude of ref.~\cite{Arkani-Hamed:2017jhn}---are included \cite{Chen:2021qkk}.
While our analysis here has been based on the Compton amplitude constructed using the minimal three-point amplitude, we have allowed for contact deformations of the same-helicity Compton amplitude as well, and fixed their contributions simultaneously to those of the opposite-helicity contact terms using \cref{eq:spinStructure}.
In all cases, we have found that this condition eliminates all contributions from the same-helicity contact terms.
We expect this to hold to all spins since at leading order in $\hbar$ all such contact terms must be proportional to $\mathfrak{a}^{4}$, and thus cannot respect \cref{eq:spinStructure} in the 2PM amplitude.
We will not say more about these contact terms below.

Where applicable, we have used Schouten identities and the vanishing of the Gram determinant to eliminate the spin structures $(\mathfrak{a}_{1}\cdot\mathfrak{a}_{2})^{2}$, $\epsilon(p_{i},q,\mathfrak{a}_{1},\mathfrak{a}_{2})$, $\epsilon(p_{1},p_{2},\mathfrak{a}_{1},\mathfrak{a}_{2})$, and $(\mathfrak{a}_{i}\cdot\mathfrak{a}_{j})\epsilon(p_{1},p_{2},q,\mathfrak{a}_{i})$ for $i\neq j$, where $\epsilon(a,b,c,d)\equiv\epsilon^{\mu\nu\alpha\beta}a_{\mu}b_{\nu}c_{\alpha}d_{\beta}$.
If $\epsilon(p_{1},p_{2},\mathfrak{a}_{1},\mathfrak{a}_{2})$ is present in the amplitude then it is not possible to impose \cref{eq:spinStructure} since terms with odd powers of $q\cdot\mathfrak{a}_{i}$ appear, which cannot be paired with an $\mathfrak{a}_{i}\cdot\mathfrak{a}_{j}$.
However, once \cref{eq:spinStructure} is imposed on the amplitude, we find that a change of basis does not affect it as long as $\epsilon(p_{1},p_{2},\mathfrak{a}_{1},\mathfrak{a}_{2})$ is not present in either basis.

The Compton amplitude in either helicity configuration cannot be deformed by contact terms at leading order in $\hbar$ up to cubic order in spin.
Computing the triangle coefficients using the Compton amplitude in \cref{eq:CorrectedGR}, we find agreement up to cubic order in spin with ref.~\cite{Chen:2021qkk}.

\subsection{Fourth order in spin}

The quartic-in-spin part of the 2PM amplitude can be divided into three independent contributions with $i$ powers of $\mathfrak{a}_{1}$ and $j$ powers of $\mathfrak{a}_{2}$ such that $i+j=4$, which we label as $(i,j)$.\footnote{The remaining $(j,i)$ contributions are related to the $(i,j)$ ones by swapping particle labels and flipping the sign of the transfer momentum.}
We find agreement with ref.~\cite{Chen:2021qkk} for the $(2,2)$ and $(3,1)$ contributions.
Considering the $(4,0)$ contribution, we have computed the amplitude including the most general contact terms that obey the black hole spin structure assumption.
We find
\begin{align}
    \cM^{(4,0)}_{2\text{PM}}&=G^{2}m_{1}^{2}m_{2}^{2}\frac{\pi^{2}}{\sqrt{-q^{2}}}\left\{M_{1}^{(4,0)}\left[(q\cdot\mathfrak{a}_{1})^{2}-q^{2}\mathfrak{a}_{1}^{2}\right]^{2}\right.\notag \\
    &\quad\left.+M_{2}^{(4,0)}q^{2}(v_{2}\cdot\mathfrak{a}_{1})^{2}\left[(q\cdot\mathfrak{a}_{1})^{2}-q^{2}\mathfrak{a}_{1}^{2}\right]+M_{3}^{(4,0)}q^{4}(v_{2}\cdot\mathfrak{a}_{1})^{4}\right\},
\end{align}
where
\begin{align*}
    M_{1}^{(4,0)}&=\frac{8m_{1}(239\omega^{4}-250\omega^{2}+35)+3m_{2}\left[64\omega^{2}(5\omega^{2}-4)+d_{0}^{(4)}(\omega^{2}-1)\right]}{768(\omega^{2}-1)}, \\
    M_{2}^{(4,0)}&=-\frac{16m_{1}(193\omega^{4}-194\omega^{2}+25)+3m_{2}\left[64(-5\omega^{4}+\omega^{2}+2)+5d_{0}^{(4)}(\omega^{2}-1)^{2}\right]}{384(\omega^{2}-1)^{2}}, \\
    M_{3}^{(4,0)}&=\frac{512m_{1}(8\omega^{4}-8\omega^{2}+1)+m_{2}\left[64(15\omega^{4}+6\omega^{2}-13)+35d_{0}^{(4)}(\omega^{2}-1)^{3}\right]}{768(\omega^{2}-1)^{3}},
\end{align*}
and $\omega\equiv v_{1}\cdot v_{2}$.
For $d_{0}^{(4)}=0$, \cref{eq:CorrectedGR} agrees with \cref{eq:HeavyCompton} at fourth order in spin, and yields equivalence at 2PM with the $(4,0)$ result in ref.~\cite{Chen:2021qkk}.
One can also see that this choice for the parameter eliminates the $\omega^{6}$ part of the polynomial in $M_{3}^{(4,0)}$, improving its behavior in the ultrarelativistic limit $\omega\rightarrow\infty$.
The result emerging from the quartic-in-spin part of the spin-5/2 Compton amplitude in refs.~\cite{Falkowski:2020aso,Chiodaroli:2021eug} corresponds to $d_{0}^{(4)}=-16/5$, while their spin-2 Compton amplitude is also described by $d_{0}^{(4)}=0$.

At leading order in $\hbar$ and this spin order, there are three contact terms that can deform the opposite-helicity Compton amplitude.
Imposing the black hole spin structure assumption fixes the coefficients of two of these to their values in \cref{eq:CorrectedGR}, and leaves the one free parameter included there.

\subsection{Fifth order in spin}

Imposing the black hole spin structure assumption uniquely fixes the $(5,0)$ part of the amplitude.
We find
\begin{align}
    \cM^{(5,0)}_{2\text{PM}}&=-iG^{2}m_{1}m_{2}\frac{\pi^{2}}{\sqrt{-q^{2}}}\omega\mathcal{E}_{1}\left\{M_{1}^{(5,0)}\left[(q\cdot\mathfrak{a}_{1})^{2}-q^{2}\mathfrak{a}_{1}^{2}\right]^{2}\right.\notag \\
    &\qquad\left.+M_{2}^{(5,0)}q^{2}(v_{2}\cdot\mathfrak{a}_{1})^{2}\left[(q\cdot\mathfrak{a}_{1})^{2}-q^{2}\mathfrak{a}_{1}^{2}\right]+M_{3}^{(5,0)}q^{4}(v_{2}\cdot\mathfrak{a}_{1})^{4}\right\},
\end{align}
where
\begin{align*}
    M_{1}^{(5,0)}&=\frac{3m_{2}(4\omega^{2}-1)+2m_{1}(13\omega^{2}-7)}{48(\omega^{2}-1)}, \\
    M_{2}^{(5,0)}&=-\frac{m_{2}(5\omega^{2}+1)+8m_{1}(2\omega^{2}-1)}{8(\omega^{2}-1)^{2}}, \\
    M_{3}^{(5,0)}&=\frac{m_{2}(-7\omega^{4}+34\omega^{2}-3)+32m_{1}(2\omega^{2}-1)}{48(\omega^{2}-1)^{3}}.
\end{align*}
We have introduced the shorthand notation $\mathcal{E}_{i}\equiv\epsilon^{\mu\nu\alpha\beta}p_{1\mu}p_{2\nu}q_{\alpha}\mathfrak{a}_{i\beta}$.

Requiring a parity-even amplitude at 2PM means there are five coefficients parametrizing possible contact term contributions to the opposite-helicity Compton amplitude.
The black hole spin structure assumption fixes two of these coefficients.
Two more are determined by applying the assumption to the $(5,1)$ part of the 2PM amplitude.
The four fixed coefficients are precisely those included in \cref{eq:CorrectedGR}, with the one remaining coefficient, $d_{0}^{(5)}$, not contributing at fifth order in spin.


\subsection{Sixth order in spin}

Moving to the next order in spin, we see the unfixed spin\textsuperscript{5} contact term coefficient entering the amplitude for the first time.
The $(6,0)$ result is
\begin{align}
    \cM^{(6,0)}_{2\text{PM}}&=G^{2}m_{1}^{2}m_{2}^{2}\frac{\pi^{2}}{\sqrt{-q^{2}}}\left\{M_{1}^{(6,0)}\left[(q\cdot\mathfrak{a}_{1})^{2}-q^{2}\mathfrak{a}_{1}^{2}\right]^{3}\right.\notag \\
    &\quad+M_{2}^{(6,0)}q^{2}(v_{2}\cdot\mathfrak{a}_{1})^{2}\left[(q\cdot\mathfrak{a}_{1})^{2}-q^{2}\mathfrak{a}_{1}^{2}\right]^{2}+M_{3}^{(6,0)}q^{4}(v_{2}\cdot\mathfrak{a}_{1})^{4}\left[(q\cdot\mathfrak{a}_{1})^{2}-q^{2}\mathfrak{a}_{1}^{2}\right]\notag \\
    &\quad\left.+M_{4}^{(6,0)}q^{6}(v_{2}\cdot\mathfrak{a}_{1})^{6}\right\},
\end{align}
where
\begin{align*}
    M_{1}^{(6,0)}&=\frac{1}{{46080(\omega^{2}-1)}}\left\{30m_{1}(149\omega^{4}-154\omega^{2}+21)\right.\notag \\
    &\quad\left.+m_{2}\left[96(20\omega^{4}-12\omega^{2}-3)+\left(d_{0}^{(4+5+6)}-5d_{1}^{(6)}\right)(\omega^{2}-1)\right]\right\}, \\
    M_{2}^{(6,0)}&=-\frac{4}{3072(\omega^{2}-1)^{2}}\left\{4m_{1}(385\omega^{4}-386\omega^{2}+49)\right.\notag \\
    &\quad\left.+m_{2}\left[32(10\omega^{4}+7\omega^{2}-11)-\left(d_{0}^{(4+5+6)}-3d_{1}^{(6)}\right)(\omega^{2}-1)^{2}\right]\right\}, \\
    M_{3}^{(6,0)}&=\frac{1}{9216(\omega^{2}-1)^{3}}\left\{768m_{1}(8\omega^{4}-8\omega^{2}+1)\right.\notag \\
    &\quad\left.+m_{2}\left[32(28\omega^{6}-39\omega^{4}+102\omega^{2}-67)+7\left(d_{0}^{(4+5+6)}-d_{1}^{(6)}\right)(\omega^{2}-1)^{3}\right]\right\}, \\
    M_{4}^{(6,0)}&=-\frac{1}{15360(\omega^{2}-1)^{4}}\left\{512m_{1}(8\omega^{4}-8\omega^{2}+1)\right.\notag \\
    &\quad\left.-m_{2}\left[32(42\omega^{8}-161\omega^{6}+201\omega^{4}-159\omega^{2}+61)+7\left(d_{0}^{(4+5+6)}+d_{1}^{(6)}\right)(\omega^{2}-1)^{4}\right]\right\},
\end{align*}
where $d_{0}^{(4+5+6)}\equiv15d_{0}^{(4)}-6d_{0}^{(5)}+d_{0}^{(6)}$.
That there are effectively only two parameters controlling this part of the amplitude is expected.
Because we have included contact terms in the $\bar{K}_{n}$ in \cref{eq:CorrectedGR}, they are all multiplied by the overall spin exponential, mixing lower- and higher-spin contact terms.
If we had added the contact terms outside the spin exponential, there would only be $\floor{n/2}-1$ coefficients contributing to the $(n,0)$ part of the 2PM amplitude.

The ultrarelativistic limits of $M_{3,4}^{(6,0)}$ can be improved by setting $d_{0}^{(4+5+6)}=-160$ and $d_{1}^{(6)}=-32$.
No other values of these parameters appear to be notable.

The opposite-helicity Compton amplitude can be deformed by 19 contact terms at leading order in $\hbar$ and sixth order in spin.
Parity invariance in conjunction with the black hole spin structure assumption determines the coefficients of 13 of these contact terms.
Four more coefficients can be pinned down by analyzing the $(6,1)$ part of the amplitude.
The values identified for the 17 coefficients are again equal to those given in \cref{eq:CorrectedGR}, and the two remaining coefficients are indeed $d_{0,1}^{(6)}$.


\subsection{Seventh order in spin}

Just like in the $(5,0)$ case, there are no free parameters contributing to the $(7,0)$ part of the amplitude once the black hole spin structure assumption is satisfied.
The unique amplitude satisfying this correspondence is
\begin{align}
    \cM^{(7,0)}_{2\text{PM}}&=-iG^{2}m_{1}m_{2}\frac{\pi^{2}}{\sqrt{-q^{2}}}\omega\mathcal{E}_{1}\left\{M_{1}^{(7,0)}\left[(q\cdot\mathfrak{a}_{1})^{2}-q^{2}\mathfrak{a}_{1}^{2}\right]^{3}\right.\notag \\
    &\quad+M_{2}^{(7,0)}q^{2}(v_{2}\cdot\mathfrak{a}_{1})^{2}\left[(q\cdot\mathfrak{a}_{1})^{2}-q^{2}\mathfrak{a}_{1}^{2}\right]^{2}+M_{3}^{(7,0)}q^{4}(v_{2}\cdot\mathfrak{a}_{1})^{4}\left[(q\cdot\mathfrak{a}_{1})^{2}-q^{2}\mathfrak{a}_{1}^{2}\right]\notag \\
    &\quad\left.+M_{4}^{(7,0)}q^{6}(v_{2}\cdot\mathfrak{a}_{1})^{6}\right\},
\end{align}
where
\begin{align*}
    M_{1}^{(7,0)}&=\frac{6m_{2}(8\omega^{2}-1)+7m_{1}(17\omega^{2}-9)}{8064(\omega^{2}-1)}, \\
    M_{2}^{(7,0)}&=-\frac{m_{2}(5\omega^{2}+1)+8m_{1}(2\omega^{2}-1)}{192(\omega^{2}-1)^{2}}, \\
    M_{3}^{(7,0)}&=\frac{m_{2}(-7\omega^{4}+34\omega^{2}-3)+32m_{1}(2\omega^{2}-1)}{576(\omega^{2}-1)^{3}}, \\
    M_{4}^{(7,0)}&=-\frac{m_{2}(9\omega^{6}-41\omega^{4}+95\omega^{2}-15)+64m_{1}(2\omega^{2}-1)}{2880(\omega^{2}-1)^{4}}.
\end{align*}

Out of 34 possible contact term coefficients, 21 can be assigned values by imposing the black hole spin structure assumption and parity invariance on the $(7,0)$ part of the amplitude.
A further eight are fixed by the $(7,1)$ calculation.
This leaves five coefficients unaccounted for up to eighth order in spin, compared to two coefficients in \cref{eq:CorrectedGR}.
It is therefore not possible to fully compare with the coefficients of contact terms in the latter at the precision we have considered.
However, because of relations among the remaining coefficients upon imposing our constraints on the amplitude, we find that computing the $(7,0)$ and $(7,1)$ contributions with the full set of contact terms or just with those present in \cref{eq:CorrectedGR} gives the same result.
We expect three of the remaining five parameters to be fixed by considering $(7,j>1)$ scattering, which would yield agreement with the number of coefficients in \cref{eq:CorrectedGR}.

The $(8,0)$ result in the auxiliary file \texttt{2PMSpin8Aux.nb} similarly has coefficients that are not fixed by considering only eighth-order-in-spin scattering.
Just like for the spin\textsuperscript{7} contact terms, cancellations among the remaining parameters mean that the result with the most general set of contact terms after imposing the black hole spin structure assumption is equivalent to the result obtained by using \cref{eq:CorrectedGR}.

\subsection{Spinless probe in a Kerr background to all orders in spin}

Taking the mass of the spinless particle to the probe limit, $m_{2}\ll m_{1}$, we can compute the leading order in the mass ratio of the 2PM scattering amplitude to all orders in spin and for arbitrary spin orientations.
When the probe particle is spinless, contact terms from the Compton amplitude do not contribute to the probe limit, which is corroborated by simple power-counting.
Therefore, the presence of finite-size effects is controlled solely by the three-point amplitude used to construct the Compton amplitude.
As we have used the minimal three-point amplitude, the results herein describe the motion of a spinless probe in a Kerr spacetime.

The scattering of a scalar with mass $m_2$ around a Kerr black hole with mass $m_1$ in the limit $m_1 \gg m_2$ is controlled by the triangle cut with the propagator of particle 1 taken on shell:
\begin{subequations}\label{eq:ProbeCut}
\begin{align}
	\mathcal{M}^{\triangleleft}_{2\text{PM}} =& \frac{G^2 \pi^2 m_1^3 m_2^2}{\sqrt{-q^2}} \left[ 6 (5\omega^2-1)  F_{0} - \frac{1}{2} F_{1} - (8\omega^4 - 8\omega^2 + 1) \sum_{n=0}^{\infty} \frac{(-1)^{n+1} 2^{4-2n}n}{(\omega^2-1)^{n+1}}   F_n \right]
	\nonumber \\ -&
	\frac{G^2 \pi^2 m_1^2 m_2}{\sqrt{-q^2}} \omega i \mathcal{E}_{1} \left[ 8F^{\prime}_{0} + (2\omega^2 - 1) \sum_{n=0}^{\infty} \frac{(-1)^{n} 2^{4-2n}(2n+1) }{(\omega^2-1 )^{n+1}} F^{\prime}_{n}  \right] , 
\end{align}
where
\begin{align}
	F_{n} &\equiv \frac{1}{\Gamma(2n+1)}\, {}_{0}F_{1}\left(2n+1; \frac{1}{4}((q\cdot \mathfrak{a}_1)^2 - q^2 \mathfrak{a}^{2}_{1}) \right) \\ &\times
	\sum_{k=0}^{n} \binom{2n}{2k} \left[q^2 (v_2\cdot \mathfrak{a}_1)^2 \right]^{n-k} \left[ q^2 (v_2\cdot \mathfrak{a}_1)^2 - (\omega^2-1)((q\cdot \mathfrak{a}_1)^2 - q^2 \mathfrak{a}_1^2 ) \right]^{k}  ,
	 \nonumber \\
	F^{\prime}_{n} &= \frac{1}{\Gamma(2n+2)}\, {}_{0}F_{1}\left(2n+2; \frac{1}{4}\left((q\cdot \mathfrak{a}_{1})^{2} - q^2 \mathfrak{a}_{1}^{2} \right) \right)
 \\ & 
 \times \sum_{k=0}^{n} \binom{2n+1}{2k+1} \left[q^2 (v_2\cdot \mathfrak{a}_1)^2 \right]^{n-k} \left[ q^2 (v_2\cdot \mathfrak{a}_1)^2 - (\omega^2-1) ((q\cdot \mathfrak{a}_{1})^2 - q^2 \mathfrak{a}_{1}^2  ) \right]^{k} ,
 \nonumber
\end{align}
\end{subequations}
and ${}_{0}F_{1}(a;z)$ is the confluent hypergeometric function. Note that dependence on the spin of the black hole enters in a combination which respects the black hole spin structure assumption.
\Cref{eq:ProbeCut} is consistent with all $(i,0)$ amplitudes presented here and in the auxiliary file \texttt{2PMSpin8Aux.nb}.

The eikonal phase in the probe limit was presented in ref.~\cite{Bautista:2021wfy} to all orders in $G$ and $\mathfrak{a}_{1}$ in the scenario of ultrarelativistic, equatorial scattering of the probe particle.
To compare to the $\mathcal{O}(G^{2})$ part of their result, we must take the ultrarelativistic limit ($\omega\rightarrow\infty$), impose equatorial-scattering kinematics, and convert \cref{eq:ProbeCut} to an eikonal phase.
Doing so yields agreement with the expansion of eq.~(5.38) in ref.~\cite{Bautista:2021wfy} for all non-negative powers of the spin.

\section{Conclusion}\label{sec:Conclusion}

Capturing all spin-multipole moments of a classical spinning object using scattering amplitudes requires scattering particles with arbitrarily high spins.
At the 1PM level, the scattering of two Kerr black holes has been fully described using amplitudes (matching the result from a purely classical approach in ref.~\cite{Vines:2017hyw}) thanks to knowledge of the appropriate three-point amplitude for arbitrary spins \cite{Arkani-Hamed:2017jhn,Guevara:2018wpp,Guevara:2019fsj,Arkani-Hamed:2019ymq}.
Moving up to 2PM, progress has been restricted to fourth order in spin because of the absence of an opposite-helicity Compton amplitude with no spurious poles above this spin order \cite{Arkani-Hamed:2017jhn,Chen:2021qkk}.
While various approaches have been taken towards remedying higher-spin Compton amplitudes \cite{Chung:2018kqs,Falkowski:2020aso,Chiodaroli:2021eug}, discrepancies between their higher-spin amplitudes (and with \cref{eq:StandardCompton} where it is valid), among other qualities, demand further attention.
Also, while the treatments in refs.~\cite{Falkowski:2020aso,Chiodaroli:2021eug} contain an analysis of potential contact terms, they do so in the context of the high-energy limit where all momenta are larger than the mass of the massive particle.
This is different from the classical limit, where the emitted bosons are soft relative to the mass.
Working in the classical limit blinded us to the high-energy behavior of the fully-quantum amplitude, but enabled us to fill many of the remaining gaps relevant to classical scattering, partly by considering the effects of contact terms on the 2PM amplitude.

Instrumental in our analysis has been the application of the heavy on-shell variables introduced in ref.~\cite{Aoude:2020onz}.
The primary computational advantages gained by recasting the problem in these variables have been the triviality of the classical limit; the expression of the opposite-helicity Compton amplitude directly in terms of the classical spin vector (\cref{eq:HeavyCompton}); the consequent isolation of the problematic part of the amplitude for high spins; and an algorithmic approach for enumerating possible contact terms.
With these variables in hand, we have succeeded in curing the opposite-helicity Compton amplitude in the classical limit for all spins for QED, QCD, and gravity.
Notable in our results is their relatively compact nature, and that the all-spin amplitudes could be written explicitly as opposed to recursively.
Of course, the simplicity of the final results can be attributed to our focus only on the leading-in-$\hbar$ part of the amplitude.

Nevertheless, this leading-in-$\hbar$ part is all that contributes to the classical gravitational amplitude at 2PM order.
In light of \cref{eq:CorrectedGR}, there is no longer an obstacle to computing the 2PM amplitude beyond fourth order in spin.
A final question remains, however: what contact terms in the Compton amplitude describe a Kerr black hole?
While we cannot definitively answer this question within the bounds of this analysis, we have investigated the 2PM amplitude in the case where the correspondence in \cref{eq:spinStructure} holds to higher orders in spin.
This assumption is based on the observation that those two spin structures always appear in this combination for black hole scattering at lower orders in spin, and in the spinless-probe limit to all spins, and this structure is broken when non-minimal three-point and higher-curvature operators are included in the scattering.
This assumption turned out to be very powerful input, fixing the contributions from almost all possible contact terms (the uniqueness of which has been checked exhaustively up to eighth order in spin), and leaving a less-than-linear growth rate for the number of new unfixed parameters with increasing spin.
As a bonus, this assumption uniquely fixes the QCD Compton amplitude at leading order in $\hbar$.
Even should the contact terms in \cref{eq:CorrectedGR} not be phenomenologically relevant for black hole scattering, the factorizable part of the amplitude presented there provides a spurious-pole-free core, expressed in terms of the classical spin vector, around which to add contact terms.

Computing with the set of contact terms producing \cref{eq:spinStructure} in the 2PM amplitude, we fully evaluated the 2PM amplitude up to eighth order in the spin of both objects in terms of the few remaining unfixed parameters.
In \Cref{sec:2PM} we presented results up to seventh order in spin for spinning-spinless scattering, while the rest of the results have been relegated to the auxiliary file \texttt{2PMSpin8Aux.nb} for brevity.
Including all contact terms for the $(4,0)$ part of the amplitude, we found that the single parameter that is unfixed by the black hole spin structure assumption is the one that translates between the fourth-order-in-spin part of \cref{eq:StandardCompton} and that of the spin-5/2 amplitudes of refs.~\cite{Falkowski:2020aso,Chiodaroli:2021eug}.
The value corresponding to \cref{eq:StandardCompton} improves the ultra-relativistic behavior of the form factor $M_{3}^{(4,0)}$, while the value for the latter references does not appear to impart notable features to the 2PM amplitude.
Interestingly, the $(5,0)$ and $(7,0)$ parts of the amplitude were uniquely fixed by imposing \cref{eq:spinStructure}, even though unfixed contact-term coefficients remained.
Going further, one can see that $M_{2,3}^{(5,0)}$ are proportional to $M_{2,3}^{(7,0)}$ (and $M_{2}^{(5,0)}$ is proportional to an analogous form factor at cubic order in spin), suggesting a potential resummation of certain parts of the odd multipoles in the spinning-spinless amplitude.
It is a simple extension of the calculation presented here to investigate whether these features hold for higher odd powers of the spin.
Moreover, it is an automatic task to extend the 2PM results in this paper to higher orders in spin.

The three-point and Compton amplitudes are all that are needed to extract classical physics at $\mathcal{O}(G^{2})$.
Above this, one must consider also the leading-in-$\hbar$ parts of higher multiplicity amplitudes.
These objects have not yet been studied in the distinct-helicity scenario.
From the simplicity of the analysis presented here, we are optimistic that the heavy on-shell variables are ideal for studying these higher-multiplicity amplitudes, and curing presumed spurious poles arising there as well.
This would open up the possibility to push the calculation of classical scattering to higher loop and spin orders.
Another extension is to include non-minimal couplings in the three-point amplitude used to build the Compton amplitude, and to attempt to cure spurious singularities in the same way as we have done here.
In this case, it would also become necessary to treat spurious poles in the same-helicity Compton amplitude \cite{Chen:2021qkk}.
Allowing for non-minimal couplings would have the interpretation of including finite-size effects.
For consistency when treating such effects, one should a priori leave all contact-term coefficients unfixed in the evaluation of the one-loop amplitude.

The on-shell approach to fixing the Compton amplitude has proven very powerful, allowing us to explicitly fix all spins at once and providing a convenient framework for systematically enumerating contact terms.
Nevertheless, a Lagrangian understanding of the results presented here would be valuable, for example for highlighting relations between the coefficients of contact terms---both within and between different helicity configurations---not evident from our on-shell perspective.
Such a description of these amplitudes may also shed light on underlying properties of the theory producing the combination in \cref{eq:spinStructure} at one-loop level.
We have found this condition to be very constraining, and even if it turns out to not describe a Kerr black hole, it remains interesting to understand why it appears to always be possible to arrange the spin in this way, and whether it expresses some underlying symmetry of the scattering.

One remaining open problem is a full matching calculation for the Compton amplitudes presented here to a classical black hole solution.
Ideally, this would fix the few remaining parameters for the contact terms, while also verifying whether imposing \cref{eq:spinStructure} on the one-loop amplitude accurately describes the dynamics of black holes.
Such a calculation could involve matching to classical EFT descriptions such as that of refs.~\cite{Bern:2020buy,Kosmopoulos:2021zoq}, in which there may be a firmer grasp on Wilson coefficient values that describe black holes.
Another approach is that of ref.~\cite{Bautista:2021wfy}, matching results from amplitudes to solutions of the Teukolsky equation.

We have opted to not derive classical observables from our amplitudes at this point.
We will leave this for future work, where we will more closely scrutinize the effects of the black hole spin structure assumption and contact terms on the cancellation of iteration pieces of the amplitude at one loop.

\acknowledgments

Calculations at one-loop order were performed using \texttt{FeynCalc} \cite{MERTIG1991345,Shtabovenko:2016sxi,Shtabovenko:2020gxv}.
We would like to thank Lucile Cangemi, Henrik Johansson, Andr\'{e}s Luna, Paolo Pichini, and Justin Vines for enlightening discussions.
We also thank Francesco Alessio, Clifford Cheung, Paolo Di Vecchia, Alessandro Georgoudis, Alex Ochirov, Julio Parra-Martinez, and Ingrid Vazquez-Holm for conversations on related topics.
We are grateful to Zvi Bern, Dimitrios Kosmopoulos, Andr\'{e}s Luna, Radu Roiban, and Fei Teng for sharing a preliminary draft of their concurrent work \cite{Bern:2022kto} and for comments on this manuscript.
We are similarly grateful to Yilber Fabian Bautista, Alfredo Guevara, Chris Kavanagh, and Justin Vines for sharing unpublished results, 
and additionally to Justin Vines for sharing unpublished notes.
We thank Henrik Johansson for providing feedback on this manuscript.
RA's research is funded by the F.R.S-FNRS project no. 40005600.
KH is grateful to Nordita for their hospitality.
KH is supported by the Knut and Alice Wallenberg Foundation under grant KAW 2018.0116 ({\it From Scattering Amplitudes to Gravitational Waves}) and the Ragnar S\"{o}derberg Foundation (Swedish Foundations’ Starting Grant).
AH is supported by the DOE under grant no. DE- SC0011632 and by the Walter Burke Institute for Theoretical Physics.

\appendix

\section{Conventions}\label{sec:Conventions}

We list here our conventions for reference. In the Weyl basis, the Dirac gamma matrices take the explicit form
\begin{align}
    \gamma^{\mu}&=\begin{pmatrix}
    0 & (\sigma^{\mu})_{\alpha\dot{\alpha}} \\
    (\bar{\sigma}^{\mu})^{\dot{\alpha}\alpha} & 0
    \end{pmatrix},
\end{align}
where $\sigma^{\mu}=(1,\sigma^{i})$, $\bar{\sigma}^{\mu}=(1,-\sigma^{i})$, and $\sigma^{i}$ are the Pauli matrices. The gamma matrices obey the Clifford algebra $\{\gamma^{\mu},\gamma^{\nu}\}=2\eta^{\mu\nu}$. We use the mostly minus metric convention, $\eta^{\mu\nu}=\text{diag}\{+,-,-,-\}$. The fifth gamma matrix is defined as
\begin{align}
    \gamma_{5}\equiv i\gamma^{0}\gamma^{1}\gamma^{2}\gamma^{3}=\begin{pmatrix}
    -\mathbb{I} & \ 0 \\
    0 & \ \mathbb{I}
    \end{pmatrix}.
\end{align}
The generator of Lorentz transforms is
\begin{align}
    J^{\mu\nu}=\frac{i}{4}[\gamma^{\mu},\gamma^{\nu}].
\end{align}

We express massless momenta in terms of on-shell variables:
\begin{subequations}
\begin{align}
    q_{\alpha\dot{\alpha}}\equiv q^{\mu}(\sigma_{\mu})_{\alpha\dot{\alpha}}&=\lambda_{\alpha}\tilde{\lambda}_{\dot{\alpha}}\equiv|\lambda\rangle_{\alpha}[\lambda|_{\dot{\alpha}}, \\
    q^{\dot{\alpha}\alpha}\equiv q^{\mu}(\bar \sigma_{\mu})^{\dot{\alpha}\alpha}&=\tilde{\lambda}^{\dot{\alpha}}\lambda^{\alpha}\equiv|\lambda]^{\dot{\alpha}}\langle\lambda|^{\alpha}.
\end{align}
\end{subequations}
Here $\alpha$, $\dot{\alpha}$ are $SL(2,\mathbb{C})$ spinor indices. Spinor brackets are formed by contracting the spinor indices,
\begin{align}
    \langle\lambda_{1}\lambda_{2}\rangle&\equiv\langle\lambda_{1}|^{\alpha}|\lambda_{2}\rangle_{\alpha}, \\
    [\lambda_{1}\lambda_{2}]&\equiv[\lambda_{1}|_{\dot{\alpha}}|\lambda_{2}]^{\dot{\alpha}}.
\end{align}
For massive momenta, we have that 
\begin{subequations}\label{eq:MassiveMom}
\begin{align}
    p_{\alpha\dot{\alpha}}&={\lambda_{\alpha}}^{I}\tilde{\lambda}_{\dot{\alpha}I}\equiv|\lambda\rangle_{\alpha}^{I}[\lambda|_{\dot{\alpha}I}, \\
    p^{\dot{\alpha}\alpha}&={\tilde{\lambda}^{\dot{\alpha}}}_{I}\lambda^{\alpha I}\equiv|\lambda]^{\dot{\alpha}}_{I}\langle\lambda|^{\alpha I},
\end{align}
\end{subequations}
where $I$ is an $SU(2)$ little group index. Spinor brackets for massive momenta are also formed by contracting spinor indices, identically to the massless case.

The Levi-Civita symbol, used to raise and lower spinor and $SU(2)$ little group indices, is defined by
\begin{align}
    \epsilon^{12}=-\epsilon_{12}=1.
\end{align}
Spinor and $SU(2)$ indices are raised and lowered by contracting with the second index on the Levi-Civita symbol. For example,
\begin{align}
    \lambda^{I}=\epsilon^{IJ}\lambda_{J},&\quad \lambda_{I}=\epsilon_{IJ}\lambda^{J}.
\end{align}

The on-shell conditions for the massive helicity variables are
\begin{subequations}\label{eq:OnShellness}
\begin{align}
    \lambda^{\alpha I}\lambda_{\alpha J}=m{\delta^{I}}_{J},\quad \lambda^{\alpha I}{\lambda_{\alpha}}^{J}&=-m\epsilon^{IJ},\quad {\lambda^{\alpha}}_{I}\lambda_{\alpha J}=m\epsilon_{IJ}, \\
    {\tilde{\lambda}_{\dot{\alpha}}}^{I}{\tilde{\lambda}^{\dot{\alpha}}}_{J}=-m{\delta^{I}}_{J},\quad {\tilde{\lambda}_{\dot{\alpha}}}^{I}\tilde{\lambda}^{\dot{\alpha} J}&=m\epsilon^{IJ},\quad \tilde{\lambda}_{\dot{\alpha} I}{\tilde{\lambda}^{\dot{\alpha}}}_{J}=-m\epsilon_{IJ}.
\end{align}
\end{subequations}
The heavy on-shell variables satisfy on-shell relations analogous to \cref{eq:OnShellness}:
\begin{subequations}\label{eq:HQETOnShellness}
\begin{align}
    \lambda^{\alpha I}_{v}\lambda_{v\alpha J}=m_{k}{\delta^{I}}_{J},\quad \lambda^{\alpha I}_{v}{\lambda_{v\alpha}}^{J}&=-m_{k}\epsilon^{IJ},\quad \lambda^{\alpha}_{vI}\lambda_{v\alpha J}=m_{k}\epsilon_{IJ}, \\
    \tilde{\lambda}_{v\dot{\alpha}}^{I}\tilde{\lambda}_{vJ}^{\dot{\alpha}}=-m_{k}{\delta^{I}}_{J},\quad \tilde{\lambda}_{v\dot{\alpha}}^{I}\tilde{\lambda}_{v}^{\dot{\alpha}J}&=m_{k}\epsilon^{IJ},\quad \tilde{\lambda}_{v\dot{\alpha}I}\tilde{\lambda}_{vJ}^{\dot{\alpha}}=-m_{k}\epsilon_{IJ},
\end{align}
where
\begin{align}
    m_{k}&\equiv\left(1-\frac{k^{2}}{4m^{2}}\right)m.
\end{align}
\end{subequations}

On-shell variables can be assigned to the upper and lower Weyl components of a Dirac spinor so that the spinors satisfy the Dirac equation \cite{Chung:2018kqs},
\begin{align}
    u^{I}(p)=\begin{pmatrix}
    {\lambda_{\alpha}}^{I} \\
    \tilde{\lambda}^{\dot{\alpha} I}
    \end{pmatrix},&\quad \bar{u}_{I}(p)=\begin{pmatrix}
    -{\lambda^{\alpha}}_{I} &    \tilde{\lambda}_{\dot{\alpha} I}
    \end{pmatrix},
\end{align}
where $p$ is expressed in terms of $\lambda$ and $\tilde{\lambda}$ as in \cref{eq:MassiveMom}.

Under a sign flip of the momentum, the on-shell variables transform as
\begin{subequations}\label{eq:AnalCont}
\begin{align}
    |-\boldsymbol{p}\rangle=-|\boldsymbol{p}\rangle,&\quad |-\boldsymbol{p}]=|\boldsymbol{p}],
\end{align}
%
which implies
\begin{align}
    |-\boldsymbol{v}\rangle = -|\boldsymbol{v}\rangle,
    &\quad |-\boldsymbol{v}]=|\boldsymbol{v}].
\end{align}
\end{subequations}

\section{Black hole spin structure assumption}\label{app:spinStructure}

We elaborate on why the groupings in \cref{eq:s1s2} are precisely those that will obey the black hole spin structure assumption in the 2PM amplitude.
To do this, we focus on the kinematics of the triangle cut, which is the cut that is relevant for the classical part of the 2PM amplitude. 

For the cuts in \cref{eq:trianglecut}, a solution for the loop momentum is
\begin{align}
	l^{\mu}_{\pm}(t) =& x q^{\mu} + y p^{\mu}_{1} + t r^{\mu}_{1} + \frac{\alpha}{t} r^{\mu}_{2} , \\
	x = \frac{2m^2_1}{q^2 - 4m^2_1}, \qquad &
	y = \frac{q^{2}}{q^2 - 4m^2_1}, \qquad
	\alpha = \frac{m^2_1 q^2}{2(q^2 - 4m^2_1) (r_1\cdot r_2)} ,
\end{align}
where $r_{1/2}$ can be written as
\begin{align}
        r^{\mu}_{1,\pm} =& \langle Q_{\pm} | \sigma^{\mu} | P_{\pm} ] , \qquad
        r^{\mu}_{2,\pm} = \langle P_{\pm} | \sigma^{\mu} | Q_{\pm} ] ,
\end{align}
with
\begin{align}
        P^{\mu}_{\pm} = p^{\mu}_{1} + \frac{m^{2}_{1}}{\gamma_{\pm}} q^{\mu} , \qquad
        Q^{\mu}_{\pm} = q^{\mu} + \frac{q^{2}}{\gamma_{\pm}} p^{\mu}_{1}, \qquad 
        \gamma_{\pm} = \frac{1}{2} \left(q^{2} \pm \sqrt{q^{2} (q^{2} - 4m^{2}_{1})} \right) .
\end{align}
The $P^{\mu}$ and $Q^{\mu}$ are null, making $r_{1/2}$ also null (and orthogonal to $p_1$ and $q$). Lastly,
\begin{align}
	\label{eq:r1r2}
        r^{\mu}_{1} r^{\nu}_{2} = 2( Q^{\mu}_{\pm}P^{\nu}_{\pm} + Q^{\nu}_{\pm} P^{\mu}_{\pm} - \eta^{\mu\nu} Q_{\pm}\cdot P_{\pm}) - 2 i \varepsilon^{\mu\nu\alpha\beta} Q_{\alpha\pm} P_{\beta\pm} .
\end{align}
We can simplify the solution to the loop momentum in the classical limit.
In the classical limit, the factors in the loop-momentum solution simplify to
\begin{align}
	x = - \frac{1}{2} + \mathcal{O}(\hbar^2) , \qquad &
	y = - \frac{q^{2}}{4m^2_1} + \mathcal{O}(\hbar^4) , \qquad
	\alpha = -\frac{q^2}{8 (r_1\cdot r_2)} + \mathcal{O}(\hbar^3) .
\end{align}
Also, since $Q^{\mu}\sim\hbar$, we have that $r_{1/2} \sim \sqrt{\hbar}$.
In the classical limit the loop momentum scales as $\mathcal{O}(\hbar)$---according to the $\hbar$ counting scheme of \cite{Kosower:2018adc}, which we have employed throughout---meaning that we should consider the scaling $t\sim\sqrt{\hbar}$.
All these elements considered, since $y p^{\mu}_1 \sim \hbar^2$, we can neglect it in the classical limit as it is subleading.

In summary, the loop-momentum solution in the classical limit is
\begin{align}
	\label{eq:loopSolClassical}
	l^{\mu}_{\pm}(t) =& -\frac{1}{2} q^{\mu} + t r^{\mu}_{1} + \frac{\alpha}{t} r^{\mu}_{2} .
\end{align}
Now we can see what happens to the spin structures in \cref{eq:s1s2} upon replacing $q_{3,4}$ with the momenta of gravitons in the loop:
\begin{align}
	\mathfrak{s}_{1} = (q_3 - q_4) \cdot \mathfrak{a}
	\rightarrow (2 l + q) \cdot \mathfrak{a}_{i} = 2 \left( t r_1 + \frac{\alpha}{t} r_2 \right)\cdot \mathfrak{a}_{i} .
\end{align}
The factors of $q\cdot \mathfrak{a}$ cancel in the classical limit in \cref{eq:loopSolClassical}.
The orthogonal combination of momenta will produce a lone $q\cdot \mathfrak{a}$ factor,
\begin{align}
	(q_3 + q_4) \cdot \mathfrak{a} \rightarrow - q\cdot \mathfrak{a}_i .
\end{align}
The other spin structure groups the spin in a similar way to $s_1$;
\begin{align}
	\mathfrak{s}_{2} = - 4 (q_3 \cdot \mathfrak{a})(q_4 \cdot \mathfrak{a}) + s_{34} \mathfrak{a}^2 \rightarrow  \mathfrak{s}^2_1 - (q\cdot \mathfrak{a}_i)^2 + q^2 (\mathfrak{a}_{i})^2 .
\end{align}
Any lone factors of $(\mathfrak{a}_{i})^2$ or $(q\cdot \mathfrak{a}_i)$ will potentially violate the black hole spin structure assumption. The way out is to group them as powers of $(q\cdot \mathfrak{a}_i)^2 - q^2 (\mathfrak{a}_{i})^2$, which is nothing but combinations of $\mathfrak{s}_1$ and $\mathfrak{s}_{2}$.

The last part of the argument is to show that combinations of $\alpha (r_1 \cdot t_1)(r_2 \cdot t_2)$ for $t_1,t_2 = p_2,\mathfrak{a}_{1},\mathfrak{a_2}$ always produce terms that respect the black hole spin structure assumption. We start with the simple cases where $t_1 = t_2$ and use \cref{eq:r1r2} to find that 
\begin{align}
	\alpha (r_1 \cdot p_2)(r_2 \cdot p_2) =& \frac{q^2 m^2_2}{16} (\omega^2 - 1) , \\
	\alpha (r_1 \cdot \mathfrak{a}_{1})(r_2 \cdot \mathfrak{a}_{1}) =& \frac{1}{16} \left[ (q\cdot \mathfrak{a}_{1})^2 - q^2 \mathfrak{a}_{1}^2 \right] , \\
	\alpha (r_1 \cdot \mathfrak{a}_{2})(r_2 \cdot \mathfrak{a}_{2}) =& \frac{1}{16} \left[  (q\cdot \mathfrak{a}_{2})^2 - q^2 \mathfrak{a}_{2}^2 + q^2 (v_1 \cdot \mathfrak{a}_{2})^2 \right] .
\end{align}
Moving to the other cases, we need to consider $t_{1}\neq t_{2}$,
\begin{align}
        \alpha (r_1 \cdot t_1) (r_2 \cdot t_2) &= \frac{1}{16} \left[ q^2(v_1 \cdot t_1)(v_1 \cdot t_2) + (q\cdot t_1) (q\cdot t_2) - q^2 (t_1 \cdot t_2) \right] - 4 \alpha i \varepsilon(t_1 t_2 q p_1 ) \nonumber \\ &
        = A(t_1,t_2) - 4 \alpha i \varepsilon(t_1 t_2 q p_1 ) ,
\end{align}
where
\begin{align}
        A(t_1,t_2) &= \frac{1}{16} \left[ q^2(v_1 \cdot t_1)(v_1 \cdot t_2) + (q\cdot t_1) (q\cdot t_2) - q^2 (t_1 \cdot t_2) \right] .
\end{align}
Calculating all combinations, we have that
\begin{align}
	A(p_2,\mathfrak{a}_{1}) &= A(\mathfrak{a}_{1},p_2) = - \frac{m_2}{16} q^2 (v_2\cdot \mathfrak{a}_{1}) , \\
	A(p_2,\mathfrak{a}_{2}) &= A(\mathfrak{a}_{2},p_2) = \frac{m_2}{16} \omega q^2 (v_1\cdot \mathfrak{a}_{2}) , \\
	A(\mathfrak{a}_{1},\mathfrak{a}_{2}) &= A(\mathfrak{a}_{2},\mathfrak{a}_{1}) = \frac{1}{16} \left[ (q\cdot \mathfrak{a}_{1})(q\cdot \mathfrak{a}_{2}) - q^2 (\mathfrak{a}_{1} \cdot \mathfrak{a}_{2}) \right] .
\end{align}
It's also easy to see by direct evaluation that products of $ \varepsilon(t_1 t_2 q p_1 ) \times \varepsilon(t^{\prime}_1 t^{\prime}_2 q p_1 )$ also only produce terms which respect the black hole spin structure assumption. Thus, all terms coming from combinations of $\alpha (r_1 \cdot t_1)(r_2 \cdot t_2)$ will respect the black hole spin structure assumption. This is sufficient to restrict the possible contact terms---formed by composing all possible combinations of \cref{eq:ContactLG,eq:ContactMonomial} for gauge theory, and their analogs for gravity---to the more manageable sets in the Compton amplitudes in \Cref{sec:HeavyCompton}.

\bibliographystyle{JHEP}
\bibliography{HeavyCompton}

\end{document}